\documentclass[10pt]{article}
\usepackage{geometry}
\usepackage{titling}
\usepackage{authblk}
\usepackage{amssymb}
\usepackage{amsmath}
\usepackage{amsthm}
\allowdisplaybreaks[4]
\usepackage{slashed}
\usepackage{graphicx,color,epsfig}
\usepackage{cite}
\usepackage[colorlinks=true,linkcolor=red,citecolor=blue]{hyperref}

\definecolor{Red}{rgb}{1.,0.,0.}
\definecolor{Blue}{rgb}{0.,0.,1.}
\newcommand{\Red}[1]{{\color{nicered}{#1}}}

\definecolor{nicered}{rgb}{0.7,0.1,0.2}
\definecolor{nicegreen}{rgb}{0.1,0.4,0.2}

\begin{document}
\title{Calculation of the $B\to K_{0,2}^*(1430)f_0(980)/\sigma$ decays in the Perturbative QCD Approach}
\author[1]{Qi-Xin Li}
\author[1]{Lei Yang}
\author[1]{Zhi-Tian Zou$\footnote{zouzt@ytu.edu.cn}$}
\author[1,2]{Ying Li$\footnote{liying@ytu.edu.cn}$}
\author[3]{Xin Liu}

\affil[1]{\it Department of Physics, Yantai University, Yantai 264005, China}
\affil[2]{\it Center for High Energy Physics, Peking University, Beijing 100871, China}
\affil[3]{\it Department of Physics, Jiangsu Normal University, Xuzhou 221116, China}
\maketitle
\vspace{0.2cm}

\begin{abstract}
Motivated by the observations of the decays $B^0 \to K_0^{*}(1430)^0 f_0(980)$ and $ B^0 \to K_2^{*}(1430)^0 f_0(980)$ from BaBar collaboration, we study the $B^{0(+)} \to K_{0,2}^{*}(1430)^{0(+)} f_0(980)/\sigma$ decays in the perturbative QCD approach for the first time. In the absence of reliable nonperturbative wave functions we only assume the scalar meson $f_0(980)$ and $\sigma$ are two-quark ground states. In our calculations, these decays are all dominated by the hard-scattering emission and annihilation diagrams, while the factorizable emission diagrams are forbidden or suppressed heavily by the vector decay constants. Furthermore, the branching fractions are sensitive to the mixing between $f_0(980)$ and $\sigma$. Comparing our results with the experimental data, a large mixing angle $\theta$ is favored. Taking $\theta=145^\circ$, the orders of branching fractions of $B \to K_0^{*}(1430)^0 \sigma$, $B \to K_{2}^{*}(1430)^0 \sigma$ and $B \to K_{0,2}^{*}(1430)^0 f_0(980)$ are predicted to be $10^{-4}$, $10^{-5}$ and $10^{-6}$, respectively, which can be measured in the current experiments such as LHCb and Belle-2. In addition, although these decays are penguin dominant, the mixing also leads to large direct $CP$ asymmetries in these decays. With the precise data in future, our results could shed light on the inner structure of the scalar mesons and can be used to determine the mixing angle of the $\sigma-f_0(980)$ system.
\end{abstract}
\newpage
\section{Introduction}
The rare $B$ meson decays have been viewed as an important place for testing the standard model \cite{Cheng:2009xz} and searching for the possible effects of new physics beyond the standard model \cite{Li:2018lxi}. In past few years, much attentions had been paid on the $B\to PP, PV$ and $VV$ decays, where $P$ and $V$ are pseudoscalar and vector mesons. With the development of high energy and high luminosity experiments, the studies of $B$ decays with scalar, axial vector and tensor particles became available.

In 2002, the decay $B\to f_0(980)K$ with large branching fraction was firstly observed in Belle experiment \cite{Abe:2002av}, and was confirmed subsequently by BaBar \cite{Aubert:2003mi} in 2004. Since then, more and more $B$ decays involving a light scalar meson in final states have been observed in both Belle \cite{Bondar:2004wr, Garmash:2004wa ,Abe:2005ig, Abe:2005nya} and BaBar \cite{Aubert:2004am, Aubert:2004hs, Aubert:2004xg, Aubert:2004fn, Aubert:2004bt, Aubert:2005ce, Aubert:2005sk, Aubert:2005wb} experiments, which provided us another perspective for the study of the scalar mesons, since their underlying structure have not been established well by studying their decays. In the theoretical side, it is well accepted by most of us that the scalar below or near 1 GeV including $\sigma$, $\kappa$, $a_0(980)$ and $f_0(980)$, form one SU(3)nonet, while the $a_0(1450)$, $K_0^*(1430)$, $f_0(1370)$, and $f_0(1500)/f_0(1710)$ with the mass above 1 GeV are grouped into another SU(3) nonet, though there is controversy around this classification. The following question is how to understand and differentiate these two nonets. For this purpose, on the basis of answering which nonet is the lowest two-quark states, two scenarios have been proposed \cite{Jaffe:1976ig,Alford:2000mm}. In the first scenario (S1), the mesons below or near 1 GeV are treated as the lowest $q\bar{q}$ bound states, and those above 1 GeV are the first excited two-quark states. On the contrary, in the another scenario (S2), the mesons near 1.5 GeV are viewed as the ground two-quark states, while the lighter mesons are identified as the predominant $qq\bar{q}\bar{q}$ states with a possible mixing with glueball states. For instance, $f_0(980)$ is the lowest two-quark state in S1, while it is a four-quark state in S2. Similarly, the heavy  scalar $K_0^*(1430)$ is the excited two quark state in S1, and in S2 it is viewed as the ground state. Of course, each scenario has its own physical picture. Taking $B$ decays with  $f_0(980)$ as an example, in S2 the light energetic $f_0(980)$ dominated by four-quark configuration requires to pick up the energetic quark-antiquark pair to form a fast four-quark state, which means that a wave function describing the interactions among four quarks are needed in the theoretical calculations \cite{Cheng:2005nb}. However, the reliable four-quark wave functions of scalar mesons are still absent till now. Therefore, we will study some particular decays in the two-quark assumption in this work. By comparing to experimental data, we hope that our results based on two-quark picture could shed light on the inner structure and characters of the scalar mesons.

In S1,the lighter scalars are regarded as the ground two-quark states. Because the $f_0(980)$ \footnote{For the sake of simplicity, we ignore the (980) and (1430) in the following context unless special statement.} is the heaviest and the $\sigma$ is the lightest one, the ideal mixing is usually adopted, and is also supported by the measurements of $D_s^+ \to f_0\pi$ and $\phi \to f_0\gamma$, which illustrates that the $f_0$ is the pure $s\bar{s}$ state. However, the observed relation $\Gamma(J/\psi\to f_0\omega) \simeq \frac{1}{2}\Gamma(J/\psi \to f_0\phi)$ \cite{Tanabashi:2018oca} implies that $f_0$ has $u\bar u$ and $d\bar d$ components. Moreover, the width of $f_0$ is dominated by the $\pi\pi$ mode, which is very similar to the case of $a_0(980)$. All the above phenomena suggest that in two-quark picture the $\sigma$ and $f_0$ should be the mixing states of $n \bar n$ and $s\bar s$  with $n\bar{n}=\frac{1}{\sqrt{2}}(u\bar{u}+d\bar{d})$, and the mixing matrix can be defined as
\begin{eqnarray}
\left(
  \begin{array}{c}
    \sigma    \\
    f_0  \\
  \end{array}
\right)
=\left(
  \begin{array}{cc}
  \cos\theta & -\sin \theta  \\
  \sin \theta& \cos\theta \\
  \end{array}
\right)\left(
         \begin{array}{c}
           n\bar n \\
           s\bar s \\
         \end{array}
       \right)\label{mixing}
\end{eqnarray}
For the $\sigma-f_0$ mixing angle $\theta$, it can be constrained by the existed experimental data. For example, using the ratio between the branching fractions of $J/\psi \to f_0\omega$ and that of $J/\psi \to f_0\phi$, the mixing angle can be obtained to be $(34\pm 6)^{\circ} \bigcup(146\pm6)^{\circ}$ \cite{Anikina:2000tj}. In ref.\cite{Cheng:2002ai}, based on the measurements of the ratio of the coupling of $f_0$ decaying into $\pi\pi$ and $KK$ the authors obtained the mixing angle to be $(25.1\pm 0.5)^{\circ} \bigcup (164.3\pm0.2)^{\circ}$ with data \cite{Aloisio:2002bt, Achasov:2000ym, Akhmetshin:1999di}, and $(42.3_{-5.5}^{+8.3})^{\circ}  \bigcup (158\pm2)^{\circ}$ with data \cite{Barberis:1999cq}. In addition, the phenomenological analysis of the radiative decay $\phi \to f_0\gamma$ and $f_0\to \gamma\gamma$ implied that the obtuse angle $\theta=(138\pm6)^{\circ}$ is more preferred. More detailed discussions about the mixing angle can be found in ref.\cite{Cheng:2002ai}. In short, it is still not clear whether there exists a universal mixing angle $\theta$ which accommodates simultaneously to all the experimental measurements. Conservatively, we set the mixing angle to be a free parameter in this work.

In 2012, BaBar collaboration reported their measurements on the decays $B^0 \to K_0^{*}(1430)^0 f_0(980)$ and $B^0 \to K_2^{*}(1430)^0 f_0(980)$\cite{Lees:2011dq}. It is only the scalar mesons and the tensor mesons that are involved in these decays, which are special in contrast to other decays with the pseudoscalar or the vector meson. When one scalar meson is produced in $B$ decays, its vector decay constant is about zero due to the conjugation invariance, and small values are caused by the violation of the SU(3) symmetry. Meanwhile, in terms of the lorentz invariance, the tensor meson cannot be produced through the $(V\pm A)$ and $(S\pm P)$ currents. Therefore, this kind of decays are highly suppressed or forbidden in naive factorization. So, in order to calculate these decays reliably, we should go beyond the naive factorization and evaluate the contributions from the nonfactorizable and annihilation type diagrams. In the past few years, the decays involving a scalar meson or a tensor meson in final states have been already explored in different approaches, such as the generalized factorization approach \cite{Giri:2006qk}, QCD factorization (QCDF) \cite{Cheng:2005nb,Cheng:2007st,Cheng:2010sn,Cheng:2010hn,Cheng:2010yd, Li:2011kw, Li:2013aca,Cheng:2013fba} and perturbative QCD approach (PQCD) \cite{Wang:2006ria, Shen:2006ms, Kim:2009dg,Liu:2009xm,Liu:2010zg, Liu:2010kq, Wang:2010ni, Zou:2012td, Zou:2012sx, Zou:2012sy, Zou:2013wza, Liu:2013lka, Liu:2013cvx, Zou:2016yhb, Zou:2017iau, Zou:2017yxc, Liu:2017cwl,Liu:2019ymi,Su:2019vbu}. Based on the researching achievements of processes, and stimulated by the experimental data, in this work ,we shall extend our studies to the $B^{0(+)} \to K_0^{*0(+)}f_0/\sigma$ and $B^{0(+)} \to K_2^{*0(+)}f_0/\sigma$ decays in PQCD approach, and try to provide new understanding to the mixing angle of the $\sigma-f_0$ mixing.

The outline of the present
paper is as follows. In Sec.~\ref{sec:function}, we introduce the formalism of the PQCD approach and the input quantities  relevant to this work, such as the decay constants and the wave functions with the light-cone distribution amplitudes. We will apply the PQCD factorization to study the $B^{0(+)} \to K_0^{*0(+)} f_0/\sigma$ and $B^{0(+)} \to K_2^{*0(+)} f_0/\sigma$ decays and present the analytic formulas of the decay amplitudes in Sec.~\ref{sec:amplitude}. The numerical result\Red{s} and the detailed discussions will be given in Sec.~\ref{sec:result}, and we will summarize this work in the last section.

\section{Formalism and Wave Function}\label{sec:function}
In the $B$ meson rest framework, because the $B$ meson is a heavy particle, the two daughter particles are energetic with large momenta and move fast. Because the light spectator quark in the $B$ meson is soft, so in order to form
an energetic final state, a hard gluon is needed to kick the soft spectator quark into a collinear one. As a result, the hard kernel is a six-quark interaction. The intrinsic character of the PQCD approach is keeping the transverse momentum $k_T$ of the valence quarks of the hadrons in the initial and final states. After that, the end-point singularity in the amplitudes will be killed naturally. Moreover, the kept transverse momenta will introduce the additional energy scale, which will lead to the double logarithms in the QCD corrections. Within the resummation technology, these double logarithms will be resumed into the so-called Sudakov form factor, which can effectively suppress the contributions from long distance.

As we have already known, there are many scales in the nonleptonic two-body $B$ meson decays, and the factorization is usually adopted. In particular, when the scale is higher than the $W$ boson mass ($m_W$), the physics can be calculated perturbatively and get the Wilson coefficients $C(m_W)$ at the scale $m_W$. Using the renormalization group, we can get the Wilson coefficients containing the physics between the scale $m_W$ and the $b$-quark mass scale ($m_b$). The physics between the scale $m_b$ and the factorization scale $t$ can be calculated perturbatively and included in the so-called hard kernel in the PQCD approach. Finally, the physics below the scale $t$ is soft and nonperturbative, which can be parameterized into the universal hadronic wave functions of the initial and final states. In this way, the decay amplitude in the PQCD approach can be written as the convolution of the Wilson coefficients $C(t)$, the hard kernel $H(x_i,b_i,t)$, and the initial and final hadronic wave functions \cite{Chang:1996dw, Yeh:1997rq}:
\begin{eqnarray}
\mathcal{A}&=&\int_0^1dx_1dx_2dx_3\int_0^{\infty}b_1db_1b_2db_2b_3db_3\,\mathbf{Tr}\Big[C(t)H(x_i,b_i,t)\nonumber\\
&& \times \Phi_B(x_1,b_1)\Phi_2(x_2,b_2)\Phi_3(x_3,b_3)S_t(x_i)e^{-S(t)}\Big],
\end{eqnarray}
$x_i(i=1,2,3)$ denoting the momentum fraction of valence quark in the meson. The $b_i$ is the conjugate variable of the transverse momentum $k_T$. The jet function $S_t(x_i)$ that is resulted from the resummation of the double logarithm $\ln^2 x_i$ can smear the end-point singularity in $x_i$ threshold effectively. The aforementioned Sudakov form factor $e^{-S(t)}$ arising from the resummation of the double logarithms $\ln^2{k_T}$ suppresses the soft dynamics effectively i.e., the long distance contributions in the large $b$ region \cite{Keum:2000ph, Li:2001ay, Keum:2000wi, Lu:2000em}. The mode-dependent hard kernel $H(x_i,b_i,t)$ and the relevant effective Hamiltonian $\mathcal{H}_{eff}$ are similar to  $B\to PP,VV$ decays, which have been discussed in detail, for example, in refs.~\cite{Ali:2007ff, Zou:2015iwa}.

In our calculations, the most important inputs are the wave functions of hadrons. For the $B$ meson, as a heavy-light system, after neglecting the numerically suppressed lorentz structure, its wave function can be defined as
\begin{eqnarray}
\Phi_B(x_1,b_1)=\frac{i}{\sqrt{2 N_c}}(\makebox[-1.5pt][l]{/}P_B+m_B)\gamma_5\phi_B(x_1,b_1),
\end{eqnarray}
with $P_B$ denoting the momentum of $B$ meson. $\phi_B(x_1,b_1)$ is the light-cone distribution amplitude (LCDA) and can be defined as
\begin{eqnarray}
\phi_B(x_1,b_1)=N_Bx_1^2(1-x_1^2)\exp\Big[-\frac{m_B^2x_1^2}{2\omega}-\frac{\omega^2b_1^2}{2}\Big].
\end{eqnarray}
In the above equation, the normalization constant $N_B$ can be determined by the normalization condition
\begin{eqnarray}
\int_0^1 dx_1 \phi_B(x_1,b_1=0)=\frac{f_B}{2\sqrt{6}},
\end{eqnarray}
where the $f_B$ is decay constant of the $B$ meson. As usual, for the shape parameter $\omega$ in the LCDA and the $f_B$,  we take $\omega=(0.4\pm 0.04)\rm GeV$, and $f_B=(0.19\pm 0.02)\rm GeV$ \cite{Keum:2000ph, Ali:2007ff, Zou:2015iwa}.

For the scalar mesons, the two decay constants can be defined as
\begin{eqnarray}
\langle S(p)|\bar{q}\gamma_{\mu}q^{\prime}|0\rangle=f_Sp_{\mu},\;\;\langle S(p)|\bar{q}q^{\prime}|0\rangle=\bar{f}_Sm_S.
\end{eqnarray}
The vector decay constant $f_S$ and the scalar decay constant $\bar{f}_S$  can be related through the equations of motion
\begin{eqnarray}
\bar{f}_S=\mu_S{f_S}=\frac{m_Sf_S}{m_2(\mu)-m_1(\mu)},
\end{eqnarray}
where $m_S$ and $m_{1(2)}$ are the scalar meson mass and the running current quark mass, respectively. From the above equation, one can find that, compared to the scalar decay constant, the vector decay constant is highly suppressed by the tiny mass difference between the two running current quark. Furthermore, for some neutral scalar mesons, such as the considered $f_0$ and $\sigma$, their vector decay constants are zero due to the charge conjugation invariance.

Up to the twist-3, the wave function of the scalar meson can be written as \cite{Cheng:2005nb, Cheng:2007st, Cheng:2013fba}.
\begin{eqnarray}
\Phi_S(x)=\frac{i}{
\sqrt{6}}\Big[\makebox[-1.5pt][l]{/}P\phi_S(x)+m_S\phi_S^{s}(x)
+m_S(\makebox[-1.5pt][l]{/}n\makebox[-1.0pt][l]{/}v-1)\phi_S^t(x)\Big],
\end{eqnarray}
with the light-like unit vectors $n=(1,0,\textbf{0}_T)$ and $v=(0,1,\textbf{0}_T)$. Similarly, the twist-2 LCDA $\phi_S(x)$ and twist-3 LCDAs $\phi_S^{s(t)}(x)$ satisfy the normalization conditions
\begin{eqnarray}
\int_0^1 dx \phi_S(x)=f_S,\;\;\;\int_0^1 dx \phi_S^{s(t)}(x)=\bar{f}_S.
\end{eqnarray}
The twist-2 LCDA $\phi_S(x,\mu)$ can be expanded as the Gegenbauer polynomials
\begin{eqnarray}
\phi_S(x)=\frac{3}{2\sqrt{6}}x(1-x)\Big[f_S +\bar{f}_S\sum_{m=1}^{\infty}B_m C_m^{3/2}(2x-1)\Big],
\end{eqnarray}
where scale-dependent $B_m$ are the Gegenbauer moments and $C_m^{3/2}$ are the
Gegenbauer polynomials. In the case of the two twist-3 LCDAs, for simplicity,
we shall adopt
the asymptotic forms \cite{Lu:2006fr}
\begin{eqnarray}
\phi_S^s(x)=\frac{\bar{f}_S}{2\sqrt{6}},\;\;\;\phi_S^t(x)=\frac{\bar{f}_S}{2\sqrt{6}}(1-2x).
\end{eqnarray}
The explicit values of the parameters $B_m$, $f_S$, and $\bar{f}_S$ are referred to the refs.\cite{Cheng:2005nb, Cheng:2007st, Cheng:2013fba}.

In the quark model, the tensor meson with $J^{PC}=2^{++}$ has the angular momentum $L = 1$ and spin $S = 1$. Due to angular momentum conservation, the polarizations with $\lambda=\pm2$ vanish in two-body $B$ decays with one tensor meson \cite{Cheng:2010hn, Cheng:2010yd}. In this case, the wave function of the tensor meson is very similar to the vector meson, and can be defined as
\begin{eqnarray}
\Phi_T&=&\frac{1}{\sqrt{6}}[m_T\makebox[-1.5pt][l]{/}\epsilon^*_{\bullet L}\phi_T(x)+\makebox[-1.5pt][l]{/}\epsilon^*_{\bullet L}\makebox[-1.5pt][l]{/}P\phi_T^t(x)+m_T^2\frac{\epsilon_{\bullet}\cdot v}{P\cdot v}\phi_T^s(x)],\nonumber\\
\Phi_T^{\perp}&=&\frac{1}{\sqrt{6}}[m_T\makebox[-1.5pt][l]{/}\epsilon^*_{\bullet \perp}\phi_T^v(x)+\makebox[-1.5pt][l]{/}\epsilon^*_{\bullet \perp}\makebox[-1.5pt][l]{/}P\phi_T^T(x)+m_Ti\varepsilon_{\mu\nu\rho\sigma}
\gamma_5\gamma^{\mu}\epsilon_{\bullet\perp}^{\ast\nu}n^{\rho}v^{\sigma}
\phi_T^a(x)].
\end{eqnarray}
with $\varepsilon^{0123}=1$. The reduced polarization vector $\epsilon_{\bullet\mu}$ can be expressed as $\epsilon_{\bullet\mu}=\frac{\epsilon_{\mu\nu}v^{\nu}}{P\cdot v}$, where the $\epsilon_{\mu\nu}$ is the polarization tensor of the tensor meson. The expressions of the twist-2 and twist-3 LCDAs are given as
\begin{eqnarray}
\phi_T(x)=\frac{f_T}{2\sqrt{6}}\phi_{\parallel}(x),\;\;
\phi_T^t(x)=\frac{f_T^{\perp}}{2\sqrt{6}}h_{\parallel}^t(x),\nonumber\\
\phi_T^s(x)=\frac{\perp}{4\sqrt{6}}\frac{d}{dx}h_{\parallel}^s(x),\;\;
\phi_T^T(x)=\frac{f_T^{\perp}}{2\sqrt{6}}\phi_{\perp}(x),\nonumber\\
\phi_T^v(x)=\frac{f_T}{2\sqrt{6}}g_{\perp}^v(x),\;\;\phi_T^a(x)=\frac{f_T}{8\sqrt{6}}\frac{d}{dx}g_{\perp}^a(x),
\end{eqnarray}
with the auxiliary functions
\begin{eqnarray}
&&\phi_{\parallel,\perp}(x)=30x(1-x)(2x-1),\;\;g_{\perp}^v(x)=5(2x-1)^3,\nonumber\\
&&h_{\parallel}^t(x)=\frac{15}{2}(2x-1)(1-6x+6x^2),\nonumber\\
&&h_{\parallel}^s(x)=15x(1-x)(2x-1),\;\;g_{\perp}^a(x)=20x(1-x)(2x-1).
\end{eqnarray}

\section{Perturbative Calculation}\label{sec:amplitude}
In this section, we shall perform the calculation of the hard kernel $H(x_i,b_i,t)$, which depends on the specific Feynman diagram. We start from the common low energy effective hamiltonian, which are given as \cite{Buchalla:1995vs}
 \begin{eqnarray}
 {\cal H}_{eff} = \frac{G_{F}}{\sqrt{2}}
     \bigg\{ V_{ub} V_{us}^{*} \big[
     C_{1}({\mu}) O_{1}({\mu})
  +  C_{2}({\mu}) O_{2}({\mu})\Big]-V_{tb} V_{ts}^{*}{\sum\limits_{i=3}^{10}} C_{i}({\mu}) O_{i}({\mu}) \bigg\} + \mbox{H.c.} ,
 \label{eq:hamiltonian}
\end{eqnarray}
where $V_{ub,us,tb,ts}$ are Cabibbo-Kobayashi-Maskawa (CKM) matrix elements. The local four-quark operators $O_{i}$ ($i=1,...,10$) are given as:
 \begin{itemize}
 \item  current--current (tree) operators
    \begin{eqnarray}
  O_{1}=({\bar{u}}_{\alpha}b_{\beta} )_{V-A}
               ({\bar{s}}_{\beta} u_{\alpha})_{V-A},
    \ \ \ \ \ \ \ \ \
   O_{2}=({\bar{u}}_{\alpha}b_{\alpha})_{V-A}
               ({\bar{s}}_{\beta} u_{\beta} )_{V-A},
    \label{eq:operator12}
    \end{eqnarray}
     \item  QCD penguin operators
    \begin{eqnarray}
      O_{3}=({\bar{s}}_{\alpha}b_{\alpha})_{V-A}\sum\limits_{q^{\prime}}
           ({\bar{q}}^{\prime}_{\beta} q^{\prime}_{\beta} )_{V-A},
    \ \ \ \ \ \ \ \ \
    O_{4}=({\bar{s}}_{\beta} b_{\alpha})_{V-A}\sum\limits_{q^{\prime}}
           ({\bar{q}}^{\prime}_{\alpha}q^{\prime}_{\beta} )_{V-A},
    \label{eq:operator34} \\
     \!\!\!\! \!\!\!\! \!\!\!\! \!\!\!\! \!\!\!\! \!\!\!\!
    O_{5}=({\bar{s}}_{\alpha}b_{\alpha})_{V-A}\sum\limits_{q^{\prime}}
           ({\bar{q}}^{\prime}_{\beta} q^{\prime}_{\beta} )_{V+A},
    \ \ \ \ \ \ \ \ \
    O_{6}=({\bar{s}}_{\beta} b_{\alpha})_{V-A}\sum\limits_{q^{\prime}}
           ({\bar{q}}^{\prime}_{\alpha}q^{\prime}_{\beta} )_{V+A},
    \label{eq:operator56}
    \end{eqnarray}
 \item electro-weak penguin operators
    \begin{eqnarray}
     O_{7}=\frac{3}{2}({\bar{s}}_{\alpha}b_{\alpha})_{V-A}
           \sum\limits_{q^{\prime}}e_{q^{\prime}}
           ({\bar{q}}^{\prime}_{\beta} q^{\prime}_{\beta} )_{V+A},
    \ \ \ \
    O_{8}=\frac{3}{2}({\bar{s}}_{\beta} b_{\alpha})_{V-A}
           \sum\limits_{q^{\prime}}e_{q^{\prime}}
           ({\bar{q}}^{\prime}_{\alpha}q^{\prime}_{\beta} )_{V+A},
    \label{eq:operator78} \\
     O_{9}=\frac{3}{2}({\bar{s}}_{\alpha}b_{\alpha})_{V-A}
           \sum\limits_{q^{\prime}}e_{q^{\prime}}
           ({\bar{q}}^{\prime}_{\beta} q^{\prime}_{\beta} )_{V-A},
    \ \ \ \
    O_{10}=\frac{3}{2}({\bar{s}}_{\beta} b_{\alpha})_{V-A}
           \sum\limits_{q^{\prime}}e_{q^{\prime}}
           ({\bar{q}}^{\prime}_{\alpha}q^{\prime}_{\beta} )_{V-A},
    \label{eq:operator9x}
    \end{eqnarray}
\end{itemize}
where $\alpha$ and $\beta$ are color indices and $q^\prime$ are the active quarks at the scale $m_b$, i.e. $q^\prime=(u,d,s,c,b)$. The left handed current is defined as $({\bar{q}}^{\prime}_{\alpha} q^{\prime}_{\beta})_{V-A}= {\bar{q}}^{\prime}_{\alpha} \gamma_\nu (1-\gamma_5)q^{\prime}_{\beta}  $ and the right handed current $({\bar{q}}^{\prime}_{\alpha} q^{\prime}_{\beta} )_{V+A}= {\bar{q}}^{\prime}_{\alpha} \gamma_\nu (1+\gamma_5) q^{\prime}_{\beta}$.  The combinations $a_i$ of Wilson coefficients are
defined as usual~\cite{Ali:1998eb}:
\begin{eqnarray}
a_1= C_2+C_1/3, &~a_2= C_1+C_2/3, &~ a_3= C_3+C_4/3,  ~a_4=
C_4+C_3/3,~a_5= C_5+C_6/3,\nonumber \\
a_6= C_6+C_5/3, &~a_7= C_7+C_8/3, &~a_8= C_8+C_7/3,~a_9=
C_9+C_{10}/3,
 ~a_{10}= C_{10}+C_{9}/3.
\end{eqnarray}

\begin{figure}[!htb]
\begin{center}
\includegraphics[scale=0.65]{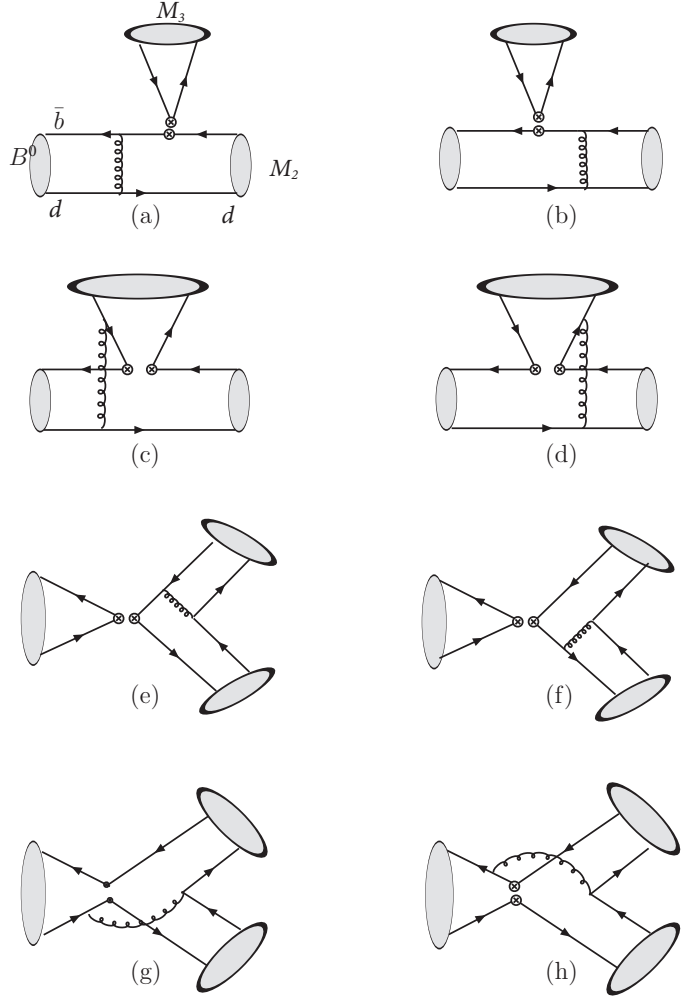}
\caption{Leading order Feynman diagrams in PQCD appraoch.}\label{Fig:1}
\end{center}
\end{figure}

In this work, we shall study two types of decays: one is the $B$ decay with two scalar mesons, while another is $B$ decay involving a tensor meson and a scalar meson. In the decay amplitudes, the subscripts $SS$ and $TS$ represent different types, respectively. According to the effective Hamiltonian (\ref{eq:hamiltonian}), we can draw the lowest order diagrams of decays we concerned, and the diagrams of decay $B \to K^{*0}_{0} f_0/\sigma$ are shown in Fig.\ref{Fig:1} as an example. These Feynman diagrams can be categorized into two classes based on the typological structures: the emission diagrams (a, b, c and d), in which the light quark in $B$ meson enter one of the light mesons as a spectator, and the annihilation diagrams (e, f, g and h), in which both of the two quarks in $B$ meson are involved in the operators. In PQCD approach, for each diagram with different operator, the whole amplitude is expressed as the convolution of the hard kernel, the related hard function, and the wave functions of involved mesons.

We first calculate the usual factorizable emission diagrams (a) and (b). When we insert the $(V-A)(V-A)$ current in the corresponding vertices, the amplitudes associated to these currents are given as:
\begin{multline}
\mathcal{F}_{SS,S}^{LL}=8\pi C_F m_B^4f_S\int_0^1dx_1dx_3\int_0^{\infty}b_1db_1b_3db_3\phi_B(x_1,b_1)\Bigg\{\Big[(1+x_3)\phi_{S3}(x_3) \\
-r_3(2x_3-1)(\phi_{S3}^s(x_3)+\phi_{S3}^t(x_3))\Big]E_{ef}(t_a)h_{ef}[x_1,x_3(1-r_2),b_1,b_3[ \\
+2r_3\phi_{S3}^s(x_3)E_{ef}(t_b)h_{ef}[x_3,x_1(1-r_2^2),b_3,b_1]\Bigg\},
\end{multline}
\begin{eqnarray}
\mathcal{F}_{TS,S}^{LL}=\sqrt{\frac{2}{3}}\mathcal{F}_{SS,S}^{LL}\mid \phi_{S3}^{(s,t)}\rightarrow \phi_T^{(s,t)},
\end{eqnarray}
where $C_F=4/3$ and $r_i=\frac{m_{M_i}}{m_B}$, with $M_i$ denoting the final states. The second term ``$S$" in the subscripts indicates that the scalar meson is emitted. The superscript ``$LL$" means the $(V-A)(V-A)$ current. The expressions of the related hard functions $E_{ef}$, $h_{ef}$, and the scale $t$ are the same as those in $B\to VV$ decays, which can be found in the Appendix of ref.\cite{Zou:2015iwa}. The $(V-A)(V+A)$ current cannot contribute to the decays we considered, so we do not include it here. When the $(S-P)(S + P)$ current, that is arising from the fierz transformation of $(V-A)(V+A)$ current, is inserted, the amplitudes can be read as
\begin{multline}
\mathcal{F}_{SS,S}^{SP}=-16 \pi C_F \bar{f}_S m_B^4 r_2\int_0^1dx_1dx_3\int_0^{\infty}b_1db_1b_3db_3\phi_B(x_1,b_1)\Bigg\{\Big[\phi_{S3}(x_3)\\
+r_3(\phi_{S3}^s(x_3)(2+x_3)-\phi_{S3}^t(x_3)x_3)\Big]E_{ef}(t_a)h_{ef}[x_1,x_3(1-r_2^2),b_1,b_3]\\
-2r_3\phi_{S3}^s(x_3)E_{ef}(t_b)h_{ef}[x_3,x_1(1-r_2^2),b_3,b_1]\Bigg\},
\end{multline}
\begin{eqnarray}
\mathcal{F}_{TS,S}^{SP}=\sqrt{\frac{2}{3}}\mathcal{F}_{SS,S}^{SP}\mid \phi_{S3}^{(s,t)}\rightarrow \phi_T^{(s,t)}.
\end{eqnarray}
Due to the fact that the tensor meson can not produced through $(V-A)$ and $(S+P)$ currents, the factorizable emission diagrams with a tensor meson emitted are forbidden, and
\begin{eqnarray}
\mathcal{F}_{TS,T}^{LL}=\mathcal{F}_{TS,T}^{SP}=0,
\end{eqnarray}

The second row in Fig.\ref{Fig:1} are the hard-scattering emission diagrams, whose decay amplitudes involve three meson wave functions. This means that the decay amplitudes are more complex than that of factorizable emission diagrams. After the variable $b_3$ is integrated out by the delta function $\delta(b_1-b_3)$, the expressions of the amplitudes are presented as follows
\begin{itemize}
  \item $(V-A)(V-A)$
  \begin{multline}
\mathcal{M}_{SS,S}^{LL}=-16\sqrt{\frac{2}{3}}C_F\pi m_B^4\int_0^1dx_1dx_2dx_3\int_0^{\infty}b_1db_1b_2db_2\phi_B(x_1,b_1)\phi_{S2}(x_2) \\
\Bigg\{\Big[\phi_{S3}(x_3)(x_2-1)+r_3x_3(\phi_{S3}^s(x_3)-\phi_{S3}^t(x_3))\Big]
E_{enf}(t_c)h_{enf}[\alpha,\beta_1,b_1,b_2] \\
+\Big[\phi_{S3}(x_3)(x_2+x_3)-r_3x_3(\phi_{S3}^s(x_3)+\phi_{S3}^t(x_3))\Big]
E_{enf}(t_d)h_{enf}[\alpha,\beta_2,b_1,b_2]\Bigg\},
\end{multline}
\begin{eqnarray}
\mathcal{M}_{TS,S}^{LL}=\sqrt{\frac{2}{3}}\mathcal{M}_{SS,S}^{LL}\mid \phi_{S3}^{(s,t)}(x_3)\rightarrow \phi_T^{(s,t)}(x_3),
\end{eqnarray}
\begin{eqnarray}
\mathcal{M}_{TS,T}^{LL}=\sqrt{\frac{2}{3}}\mathcal{M}_{SS,S}^{LL}\mid \phi_{S2}(x_2)\rightarrow \phi_T(x_2).
\end{eqnarray}
  \item $(V-A)(V+A)$
\begin{multline}
\mathcal{M}_{SS,S}^{LR}=16\sqrt{\frac{2}{3}}C_F\pi r_2m_B^4\int_0^1dx_1dx_2dx_3\int_0^{\infty}b_1db_1b2db_2\phi_B(x_1,b_1)\\
\Bigg\{\Big[(x_2-1)\phi_{S3}(x_3)(\phi_{S2}^s(x_2)+\phi_{S2}^t(x_2))
+r_3\Big((1-x_2+x_3)(\phi_{S2}^t(x_2)\phi_{S3}^t(x_3)-\phi_{S2}^s(x_2)\phi_{S3}^s(x_3))\\
+(x_2+x_3-1)(\phi_{S2}^t(x_2)\phi_{S3}^s(x_3)-\phi_{S2}^s(x_2)\phi_{S3}^t(x_3))\Big)\Big]
E_{enf}(t_c)h_{enf}[\alpha,\beta_1,b_1,b_2]\\
+\Big[x_2\phi_{S3}(x_3)(\phi_{S2}^s(x_2)-\phi_{S2}^t(x_2))
+r_3\Big((x_3-x_2)(\phi_{S2}^s(x_2)\phi_{S3}^t(x_3)+\phi_{S2}^t(x_2)\phi_{S3}^s(x_3))\\
+(x_2+x_3)(\phi_{S2}^s(x_2)\phi_{S3}^s(x_3)+\phi_{S2}^t(x_2)\phi_{S3}^t(x_3)\Big)\Big]
E_{enf}(t_d)h_{enf}[\alpha,\beta_2,b_1,b_2]\Bigg\},
\end{multline}
\begin{eqnarray}
\mathcal{M}_{TS,S}^{LR}=\sqrt{\frac{2}{3}}\mathcal{M}_{SS,S}^{LR}\mid \phi_{S3}^{(s,t)}(x_3)\rightarrow \phi_T^{(s,t)}(x_3),
\end{eqnarray}
\begin{eqnarray}
\mathcal{M}_{TS,T}^{LR}=\sqrt{\frac{2}{3}}\mathcal{M}_{SS,S}^{LR}\mid \phi_{S2}^{(s,t)}(x_2)\rightarrow \phi_T^{(s,t)}(x_2),
\end{eqnarray}
  \item $(S-P)(S+P)$
\begin{multline}
\mathcal{M}_{SS,S}^{SP}=-16\sqrt{\frac{2}{3}}C_F\pi m_B^4\int_0^1dx_1dx_2dx_3\int_0^{\infty}b_1db_1b_2db_2\phi_B(x_1,b_1)\phi_{S2}(x_2)\\
\Bigg\{\Big[\phi_{S3}(x_3)(-1+x_2-x_3)+r_3x_3(\phi_{S3}^s(x_3)+\phi_{S3}^t(x_3))\Big]
E_{enf}(t_c)h_{enf}[\alpha,\beta_1,b_1,b_2]\\
+\Big[\phi_{S3}(x_3)x_2+r_3x_3(\phi_{S3}^t(x_3)-\phi_{S3}^s(x_3))]E_{enf}(t_d)h_{enf}[\alpha,\beta_2,b_1,b_2]\Bigg\},
\end{multline}
\begin{eqnarray}
\mathcal{M}_{TS,S}^{SP}=\sqrt{\frac{2}{3}}\mathcal{M}_{SS,S}^{SP}\mid \phi_{S3}^{(s,t)}(x_3)\rightarrow \phi_T^{(s,t)}(x_3),
\end{eqnarray}
\begin{eqnarray}
\mathcal{M}_{TS,T}^{SP}=\sqrt{\frac{2}{3}}\mathcal{M}_{SS,S}^{SP}\mid \phi_{S2}(x_2)\rightarrow \phi_T(x_2).
\end{eqnarray}
\end{itemize}
Particularly, when the emitted meson is a  pseudoscalar or a vector light meson, the total contributions of these nonfactorizable emission diagrams are suppressed highly, due to the cancelation between the two diagrams (c and d). While for the current considered decays with a scalar/tensor meson emitted, because LCDAs are antisymmetric, the contributions between the two diagrams are no longer destructive but constructive. Therefore, the  nonfactorizable emission diagrams contributions are not suppressed in those considered decays.

Now we move to calculate the annihilation diagrams, where two quarks in the initial $B$ meson are involved the four-quark interaction and $q\bar q$ quarks included in  final states are produced from a hard gluon. In Figure.\ref{Fig:1}, the diagrams (e and f) in third row are the so-called factorizable annihilation type diagrams, whose decay amplitudes can be calculated as follow:
\begin{itemize}
  \item $(V-A)(V\pm A)$ current
\begin{multline}
\mathcal{A}_{SS,S}^{LL(LR)}=8C_F f_B\pi m_B^4\int_0^1dx_2dx_3\int_0^{\infty}b_2db_2b_3db_3\Bigg\{\Big[(x_3-1)\phi_{S2}(x_2)\phi_{S3}(x_3)\\
+2r_2r_3\phi_{S2}^s(x_2)(\phi_{S3}^s(x_3)(x_3-2)+\phi_{S3}^t(x_3)x_3)\Big]
E_{af}(t_e)h_{af}[\alpha_1,\beta,b_2,b_3]\\
+\Big[-2r_2r_3\phi_{S3}^s(x_3)(\phi_{S2}^s(x_2)(1+x_2)+\phi_{S2}^t(x_2)(x_2-1))\\
+x_2\phi_{S2}(x_2)\phi_{S3}(x_3)\Big]E_{af}(t_f)h_{af}[\alpha_2,\beta,b_2,b_3]\Bigg\},
\label{da:wava}
\end{multline}
\begin{eqnarray}
\mathcal{A}_{TS,S}^{LL(LR)}=\sqrt{\frac{2}{3}}\mathcal{A}_{SS,S}^{LL(LR)}|\phi_{S3}^{(s,t)}(x_3)\rightarrow \phi_T^{(s,t)}(x_3),
\end{eqnarray}
\begin{eqnarray}
\mathcal{A}_{TS,T}^{LL(LR)}=\sqrt{\frac{2}{3}}\mathcal{A}_{SS,S}^{LL(LR)}|\phi_{S2}^{(s,t)}(x_2)\rightarrow \phi_T^{(s,t)}(x_2),
\end{eqnarray}
\item $(S-P)(S+P)$ current
\begin{multline}
\mathcal{A}_{SS,S}^{SP}=-16C_Ff_B\pi m_B^4\int_0^1dx_2dx_3\int_0^{\infty}b_2db_2b_3db_3
\Bigg\{\Big[2 r_2 \phi_{S3}(x_3)\phi_{S2}^s(x_2)\\
+r_3(x_3-1)\phi_{S2}(x_2)(\phi_{S3}^s(x_3)+\phi_{S3}^t(x_3))\Big]E_{af}(t_e)h_{af}[\alpha_1,\beta,b_2,b_3]\\
-\Big[2r_3\phi_{S2}(x_2)\phi_{S3}^s(x_3)+r_2x_2\phi_{S3}(x_3)(\phi_{S2}^t(x_2)-\phi_{S2}^s(x_2))\Big]
E_{af}(t_f)h_{af}[\alpha_2,\beta,b_2,b_3]\Bigg\},
\end{multline}
\begin{eqnarray}
\mathcal{A}_{TS,S}^{SP}=\sqrt{\frac{2}{3}}\mathcal{A}_{SS,S}^{SP}|\phi_{S3}^{(s,t)}(x_3)\rightarrow \phi_T^{(s,t)}(x_3),
\end{eqnarray}
\begin{eqnarray}
\mathcal{A}_{TS,T}^{SP}=\sqrt{\frac{2}{3}}\mathcal{A}_{SS,S}^{SP}|\phi_{S2}^{(s,t)}(x_2)\rightarrow \phi_T^{(s,t)}(x_2).
\end{eqnarray}
\end{itemize}
In above equations, the related scales $t_{e,f}$, the functions $h_{af}$ and the inner functions can be found in the Appendix of ref.\cite{Zou:2015iwa}.

The amplitude for the nonfactorizable annihilation diagram in Fig.1(g) and (h) results in
\begin{itemize}
\item $(V-A)(V-A)$
\begin{multline}
\mathcal{W}_{SS,S}^{LL}=16\sqrt{\frac{2}{3}}C_F \pi m_B^4 \int_0^1dx_1dx_2dx_3\int_0^{\infty}b_1db_1b_2db_2\phi_B(x_1,b_1) \\
\Bigg\{\Big[-\phi_{S2}(x_2)\phi_{S3}(x_3)x_2
+r_2r_3\Big(\phi_{S2}^t(x_2)(\phi_{S3}^t(x_3)(1-x_2+x_3)+\phi_{S3}^s(x_3)(x_2+x_3-1))\\
+\phi_{S2}^s(x_2)(\phi_{S3}^t(x_3)(1-x_2-x_3)+\phi_{S3}^s(x_3)(3+x_2-x_3))\Big)\Big]
E_{anf}(t_g)h_{anf}[\alpha,\beta_1,b_1,b_2]\\
-\Big[\phi_{S2}(x_2)\phi_{S3}(x_3)(x_3-1)
+r_2r_3\Big(\phi_{S2}^s(x_2)(\phi_{S3}^s(x_3)(1+x_2-x_3)-\phi_{S3}^t(x_3)(1-x_2-x_3))\\
+\phi_{S2}^t(x_2)(\phi_{S3}^s(x_3)(1-x_2-x_3)
-\phi_{S3}^t(x_3)(1+x_2-x_3))\Big)\Big]E_{anf}(t_h)h_{anf}[\alpha,\beta_2,b_1,b_2]\Bigg\},
\end{multline}
\begin{eqnarray}
\mathcal{W}_{TS,S}^{LL}=\sqrt{\frac{2}{3}}\mathcal{W}_{SS,S}^{LL}|\phi_{S3}^{(s,t)}(x_3)\rightarrow \phi_T^{(s,t)}(x_3),
\end{eqnarray}
\begin{eqnarray}
\mathcal{W}_{TS,T}^{LL}=\sqrt{\frac{2}{3}}\mathcal{W}_{SS,S}^{LL}|\phi_{S2}^{(s,t)}(x_2)\rightarrow \phi_T^{(s,t)}(x_2),
\end{eqnarray}
\item $(V-A)(V+A)$
\begin{multline}
\mathcal{W}_{SS,S}^{LR}=16\sqrt{\frac{2}{3}}C_F \pi m_B^4\int_0^1dx_1dx_2dx_3\int_0^{\infty}b_1db_1b_2db_2\phi_B(x_1,b_1)\\
\Bigg\{\Big[r_2\phi_{S3}(x_3)(\phi_{S2}^s(x_2)+\phi_{S2}^t(x_2))(x_2-2)-r_3\phi_{S2}(x_2)(\phi_{S3}^s(x_3)\\
-\phi_{S3}^t(x_3))(x_3+1)\Big]E_{anf}(t_g)h_{anf}[\alpha,\beta_1,b_1,b_2]\\
+\Big[-r_2x_2\phi_{S3}(x_3)(\phi_{S2}^s(x_2)+\phi_{S2}^t(x_2))+r_3(x_3-1)\phi_{S2}(x_2)(\phi_{S3}^s(x_3)\\
-\phi_{S3}^t(x_3))\Big]E_{anf}(t_h)h_{anf}[\alpha,\beta_2,b_1,b_2]\Bigg\},
\end{multline}
\begin{eqnarray}
\mathcal{W}_{TS,S}^{LR}=\mathcal{W}_{SS,S}^{LR}|\phi_{S3}^{(s,t)}(x_3)\rightarrow \phi_T^{(s,t)}(x_3),
\end{eqnarray}
\begin{eqnarray}
\mathcal{W}_{TS,T}^{LR}=\mathcal{W}_{SS,S}^{LR}|\phi_{S2}^{(s,t)}(x_2)\rightarrow \phi_T^{(s,t)}(x_2),
\end{eqnarray}
\item $(S-P)(S+P)$
\begin{multline}
\mathcal{W}_{SS,S}^{SP}=16\sqrt{\frac{2}{3}}C_F\pi m_B^4\int_0^1dx_1dx_2dx_3\int_0^{\infty}b_1db_1b_2db_2\phi_B(x_1,b_1)\\
\Bigg\{\Big[(x_3-1)\phi_{S2}(x_2)\phi_{S3}(x_3)+r_2r_3\Big(\phi_{S2}^t(x_2)(\phi_{S3}^t(x_3)(1-x_2+x_3)
-\phi_{S3}^s(x_3)(x_2+x_3-1))\\
+\phi_{S2}^s(x_2)(\phi_{S3}^s(x_3)(3+x_2-x_3)
+\phi_{S3}^t(x_3)(x_2+x_3-1))\Big)\Big]E_{anf}(t_g)h_{anf}[\alpha,\beta_1,b_1,b_2]\\
+\Big[x_2\phi_{S2}(x_2)\phi_{S3}(x_3)-r_2r_3\Big(\phi_{S2}^s(x_2)(\phi_{S3}^s(x_3)(1+x_2-x_3)
+\phi_{S3}^t(x_3)(1-x_2-x_3))\\
+\phi_{S2}^t(x_2)(\phi_{S3}^t(x_3)(-1-x_2+x_3)
+\phi_{S3}^s(x_3)(x_2+x_3-1))\Big)\Big]E_{anf}(t_h)h_{anf}[\alpha,\beta_2,b_1,b_2]\Bigg\},
\end{multline}
\begin{eqnarray}
\mathcal{W}_{TS,S}^{SP}=\sqrt{\frac{2}{3}}\mathcal{W}_{SS,S}^{SP}|\phi_{S3}^{(s,t)}(x_3)\rightarrow \phi_T^{(s,t)}(x_3),
\end{eqnarray}
\begin{eqnarray}
\mathcal{W}_{TS,T}^{SP}=\sqrt{\frac{2}{3}}\mathcal{W}_{SS,S}^{SP}|\phi_{S2}^{(s,t)}(x_2)\rightarrow \phi_T^{(s,t)}(x_2).
\end{eqnarray}
\end{itemize}
The related functions and the scales($t_g$ and $t_h$) can be referred in the ref.\cite{Zou:2015iwa}. From the Eq.(\ref{da:wava}), it is obvious that there exist large cancellations between the two annihilation type diagrams (e and f), thus the annihilation diagrams is viewed as  power suppressed. This picture is consistent with the naive argument about the neglect of the annihilation type diagrams \cite{Wirbel:1985ji,Bauer:1986bm}. However, although these diagrams are power suppressed, they can provide a large strong phase, which is used to explain the CP asymmetry in $B$ decays \cite{Keum:2000ph,Li:2001ay,Keum:2000wi,Lu:2000em,Ali:2007ff}.

Finally, the total amplitude of $B\to K_{0}^{*+}S$ can be written as：
\begin{multline}
\mathcal{A}(B^+ \to K_{0}^{*+}S(n\bar n)) =\frac{G_F}{2}\Bigg\{V_{ub}^*V_{us}\Big[a_1(\mathcal{F}_{SS,S}^{LL}+\mathcal{A}_{SS,S}^{LL})
+C_1(\mathcal{M}_{SS,S}^{LL}+\mathcal{W}_{SS,S}^{LL})\Big]\\
-V_{tb}^*V_{ts}\Big[(a_4+a_{10})(\mathcal{F}_{SS,S}^{LL}+\mathcal{A}_{SS,S}^{LL})
+(a_6+a_8)(\mathcal{F}_{SS,S}^{SP}+\mathcal{A}_{SS,S}^{SP})\\
+(C_3+C_9)(\mathcal{M}_{SS,S}^{LL}+\mathcal{W}_{SS,S}^{LL})
+(C_5+C_7)(\mathcal{M}_{SS,S}^{LR}+\mathcal{W}_{SS,S}^{LR})\Big]\Bigg\},
\end{multline}
\begin{multline}
\mathcal{A}(B^+ \to K_{0}^{*+}S(s\bar s))=\frac{G_F}{\sqrt{2}}\Bigg\{V_{ub}^*V_{us}C_2\mathcal{M}_{SS,S}^{LL}
-V_{tb}^*V_{ts}\Big[\left(2C_4+\frac{1}{2}C_{10}\right)\mathcal{M}_{SS,S}^{LL}
+\left(2C_6+\frac{1}{2}C_8\right)\mathcal{M}_{SS,S}^{SP}\Big]\Bigg\},
\end{multline}
\begin{multline}
\mathcal{A}(B^0\to K_{0}^{*0}S(n\bar n))=\frac{G_F}{2}\Bigg\{V_{ub}^*V_{us}C_1\mathcal{M}_{SS,S}^{LL}
-V_{tb}^*V_{ts}\Big[\left(a_4-\frac{1}{2}a_{10}\right)(\mathcal{F}_{SS,S}^{LL}+\mathcal{A}_{SS,S}^{LL}) \\
+\left(a_6-\frac{1}{2}a_{8}\right)(\mathcal{F}_{SS,S}^{SP}+\mathcal{A}_{SS,S}^{SP})
+\left(C_3-\frac{1}{2}C_9\right)(\mathcal{M}_{SS,S}^{LL}+\mathcal{W}_{SS,S}^{LL})\\
+\left(C_5-\frac{1}{2}C_7\right)(\mathcal{M}_{SS,S}^{LR}+\mathcal{W}_{SS,S}^{LR})
+\left(2C_4+\frac{1}{2}C_{10}\right)\mathcal{M}_{SS,S}^{LL}+(2C_6+C_8/2)\mathcal{M}_{SS,S}^{SP}\Big]\Bigg\},
\end{multline}
\begin{multline}
\mathcal{A}(B^0\to K_{0}^{*0}S(s\bar s))= \frac{G_F}{\sqrt{2}}\Bigg\{-V_{tb}^*V_{ts}\Big[
 \left(a_6-\frac{1}{2}a_{8}\right)(\mathcal{F}_{SS,S}^{SP}+\mathcal{A}_{SS,S}^{SP})
+\left(a_4-\frac{1}{2}a_{10}\right)\mathcal{A}_{SS,S}^{LL}\\
+\left(C_3+C_4-\frac{1}{2}C_9-\frac{1}{2}C_{10}\right)\mathcal{M}_{SS,S}^{LL}
+\left(C_3-\frac{1}{2}C_9\right)\mathcal{W}_{SS,S}^{LL}\\
+\left(C_5-\frac{1}{2}C_7\right)(\mathcal{M}_{SS,S}^{LR}+\mathcal{W}_{SS,S}^{LR})
+\left(C_6-\frac{1}{2}C_8\right)\mathcal{M}_{SS,S}^{SP}
\Big]\Bigg\},
\end{multline}
\begin{multline}
\mathcal{A}(B^+ \to K_{2}^{*+}S(n\bar n)) =\frac{G_F}{2}\Bigg\{V_{ub}^*V_{us}\Big[a_1(\mathcal{F}_{SS,S}^{LL}+\mathcal{A}_{TS,T}^{LL})
+C_1(\mathcal{M}_{TS,T}^{LL}+\mathcal{W}_{TS,T}^{LL})\Big]\\
-V_{tb}^*V_{ts}\Big[(a_4+a_{10})(\mathcal{F}_{TS,T}^{LL}+\mathcal{A}_{SS,S}^{LL})
+(a_6+a_8)(\mathcal{F}_{TS,T}^{SP}+\mathcal{A}_{TS,T}^{SP})\\
+(C_3+C_9)(\mathcal{M}_{TS,T}^{LL}+\mathcal{W}_{TS,T}^{LL})
+(C_5+C_7)(\mathcal{M}_{TS,T}^{LR}+\mathcal{W}_{TS,T}^{LR})\Big]\Bigg\},
\end{multline}
\begin{multline}
\mathcal{A}(B^+ \to K_{2}^{*+}S(s\bar s))=\frac{G_F}{\sqrt{2}}\Bigg\{V_{ub}^*V_{us}C_2\mathcal{M}_{TS,S}^{LL}
-V_{tb}^*V_{ts}\Big[\left(2C_4+\frac{1}{2}C_{10}\right)\mathcal{M}_{TS,S}^{LL}
+\left(2C_6+\frac{1}{2}C_8\right)\mathcal{M}_{TS,S}^{SP}\Big]\Bigg\},
\end{multline}
\begin{multline}
\mathcal{A}(B^0\to K_{0}^{*0}S(n\bar n))=\frac{G_F}{2}\Bigg\{V_{ub}^*V_{us}C_1\mathcal{M}_{TS,T}^{LL}
-V_{tb}^*V_{ts}\Big[\left(a_4-\frac{1}{2}a_{10}\right)(\mathcal{F}_{TS,T}^{LL}+\mathcal{A}_{TS,T}^{LL}) \\
+\left(a_6-\frac{1}{2}a_{8}\right)(\mathcal{F}_{TS,T}^{SP}+\mathcal{A}_{TS,T}^{SP})
+\left(C_3-\frac{1}{2}C_9\right)(\mathcal{M}_{TS,T}^{LL}+\mathcal{W}_{TS,T}^{LL})\\
+\left(C_5-\frac{1}{2}C_7\right)(\mathcal{M}_{TS,T}^{LR}+\mathcal{W}_{TS,T}^{LR})
+\left(2C_4+\frac{1}{2}C_{10}\right)\mathcal{M}_{TS,T}^{LL}+(2C_6+C_8/2)\mathcal{M}_{TS,T}^{SP}\Big]\Bigg\},
\end{multline}
\begin{multline}
\mathcal{A}(B^0\to K_{0}^{*0}S(s\bar s))= \frac{G_F}{\sqrt{2}}\Bigg\{-V_{tb}^*V_{ts}\Big[
 \left(a_6-\frac{1}{2}a_{8}\right)(\mathcal{F}_{TS,S}^{SP}+\mathcal{A}_{TS,S}^{SP})
+\left(a_4-\frac{1}{2}a_{10}\right)\mathcal{A}_{TS,S}^{LL}\\
+\left(C_3+C_4-\frac{1}{2}C_9-\frac{1}{2}C_{10}\right)\mathcal{M}_{TS,S}^{LL}
+\left(C_3-\frac{1}{2}C_9\right)\mathcal{W}_{TS,S}^{LL}\\
+\left(C_5-\frac{1}{2}C_7\right)(\mathcal{M}_{TS,S}^{LR}+\mathcal{W}_{TS,S}^{LR})
+\left(C_6-\frac{1}{2}C_8\right)\mathcal{M}_{TS,S}^{SP}
\Big]\Bigg\}.
\end{multline}
Then, we can write down the amplitudes of $B\to K_{0,2}^{*}f_0$ and $B\to K_{0,2}^{*}\sigma$ as
\begin{eqnarray}
\mathcal{A}(B \to K_{0(2)}^{*}f_0)&=&\mathcal{A}(B \to K_{0(2)}^{*}S(n\bar n))\sin{\theta}+\mathcal{A}(B \to K_{0(2)}^{*}S(s\bar s))\cos{\theta},\\
\mathcal{A}(B \to K_{0(2)}^{*}\sigma)&=&\mathcal{A}(B \to K_{0(2)}^{*}S(n\bar n))\cos{\theta}-\mathcal{A}(B \to K_{0(2)}^{*}S(s\bar s))\sin{\theta}.
\end{eqnarray}
Meanwhile, the direct CP asymmetries of these decays can be defined as
\begin{eqnarray}
A_{CP}=\frac{\mathcal{A}(B^0\to K_{0(2)}^{*}f_0)-\mathcal{A}(\overline B^0 \to \overline K_{0(2)}^{*}f_0)}
{\mathcal{A}(B^0 \to K_{0(2)}^{*}f_0)+\mathcal{A}(\overline B^0 \to \overline K_{0(2)}^{*}f_0)},
\end{eqnarray}
Obviously, both the amplitudes and the direct CP asymmetries are related to the mixing angle $\theta$.

\section{Numerical Results and Discussions}\label{sec:result}
We start this section by setting constants used in the calculations. The vector decay constants and the scalar decay constants of the $f_0$ and $\sigma$ can be found in ref.\cite{Cheng:2013fba}. Other input parameters such as the QCD scale, the masses of
mesons, the CKM matrix elements, the decay constant of the $B$ meson and the lifetimes of the $B$ mesons (in $\rm ps$) are adopted as follows \cite{Tanabashi:2018oca}:
\begin{align}
&\Lambda_{\overline{\rm MS}}^{f=4}=0.25\pm 0.05 {~\rm GeV},\;\;
m_B=5.279 {~\rm GeV},\;\;
m_b=4.8{~\rm GeV},\;\;\nonumber\\
&\tau_{B^{\pm/0}}=1.638/1.519{~\rm ps},\;\;
V_{ub}=0.00365 ,\;\;
V_{us}=0.22452,\;\;\nonumber\\
&V_{ts}=0.04133,\;\;
V_{tb}=0.999105,\;\;f_B=0.19\pm0.02 {~\rm GeV}.
\end{align}

\begin{figure}[!htb]
\begin{center}
\includegraphics[scale=0.65]{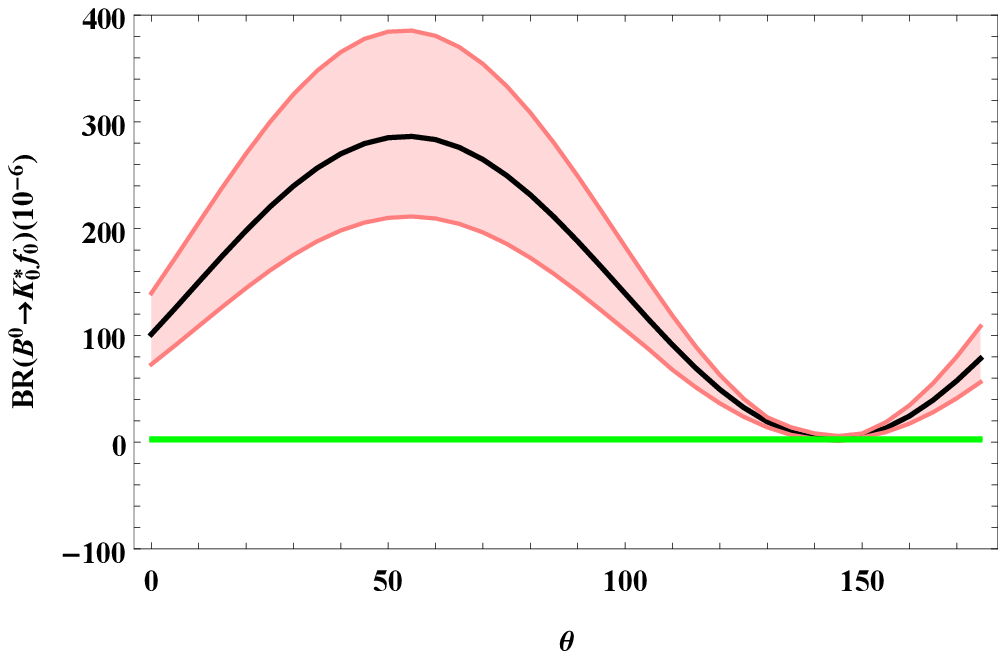}\,\,\,\,
\includegraphics[scale=0.65]{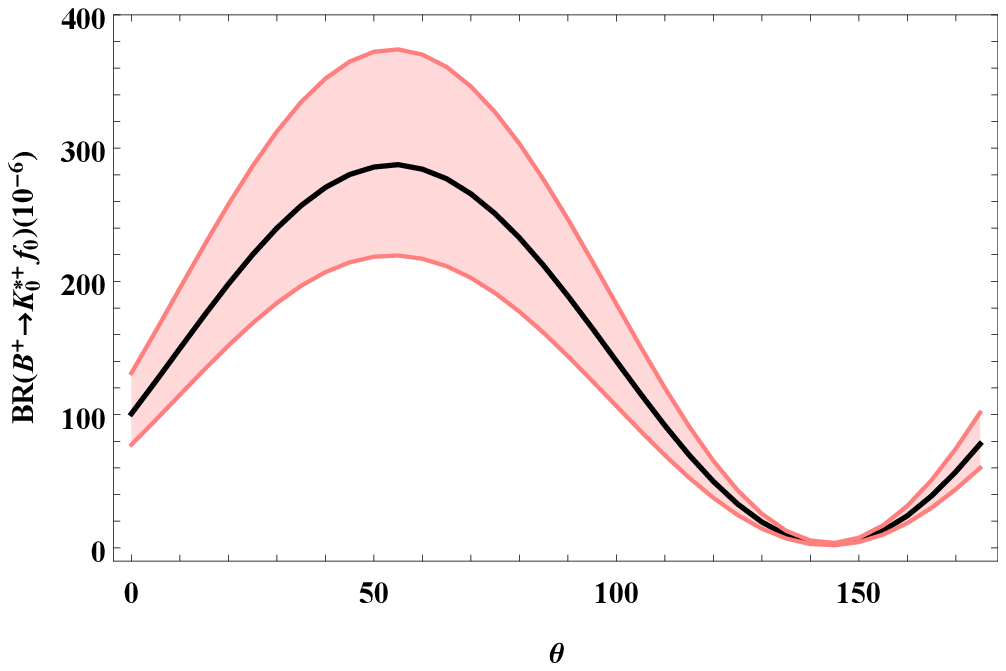}\\
\includegraphics[scale=0.65]{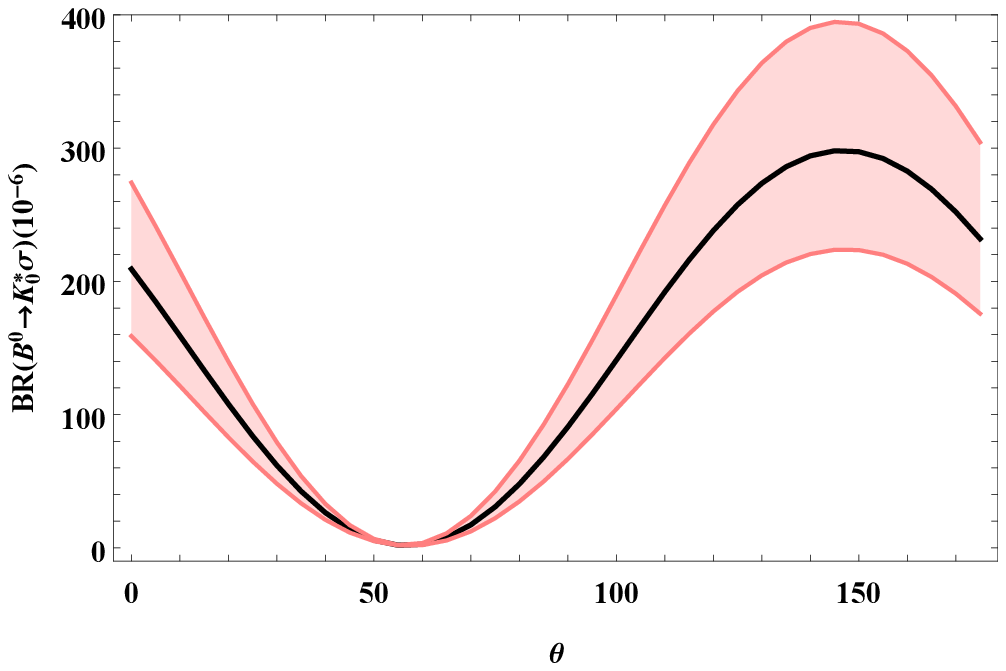}\,\,\,\,
\includegraphics[scale=0.65]{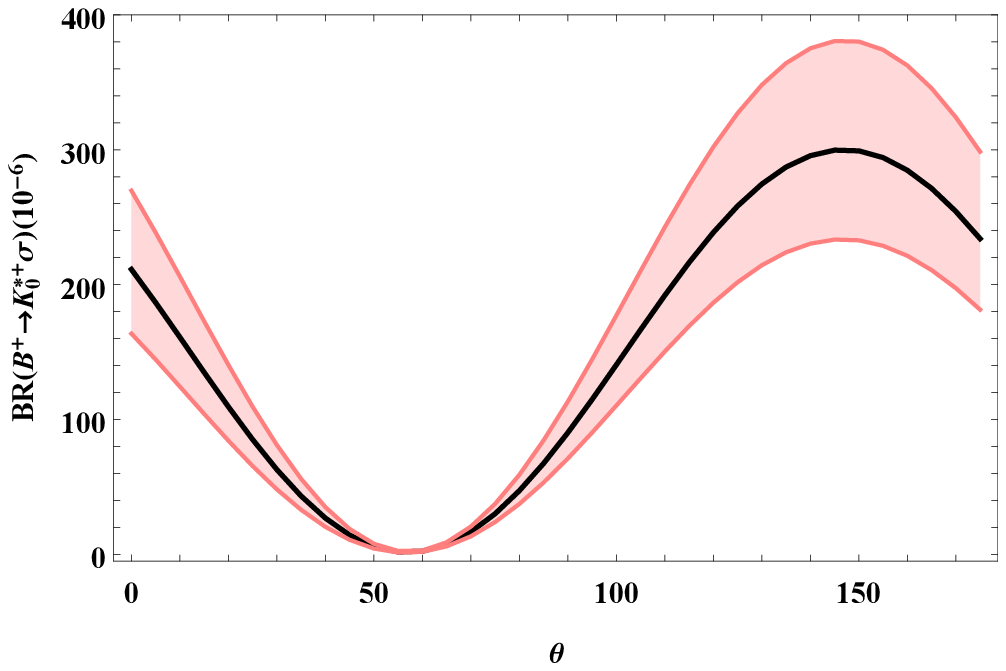}
\caption{The branching fractions of $B\to K_0^* f_0$ and $B\to K_2^* \sigma$ with variant of the mixing angle $\theta$. The black lines are the center values, and the horizontal (green) band is the experimental value.}\label{Fig:2}
\end{center}
\end{figure}

\begin{figure}[!htb]
\begin{center}
\includegraphics[scale=0.65]{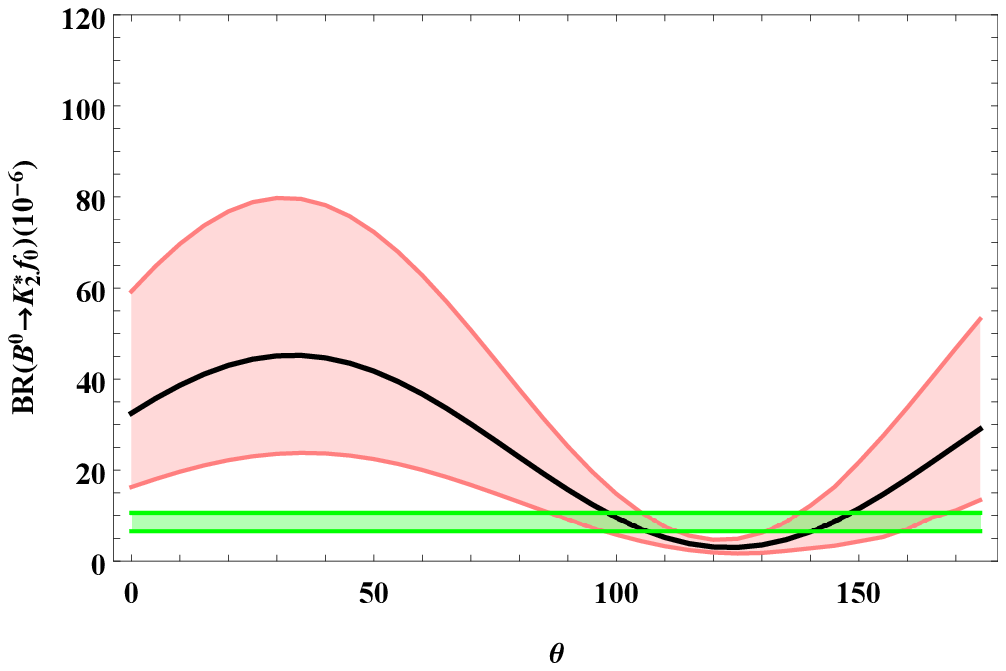}\,\,\,\,
\includegraphics[scale=0.65]{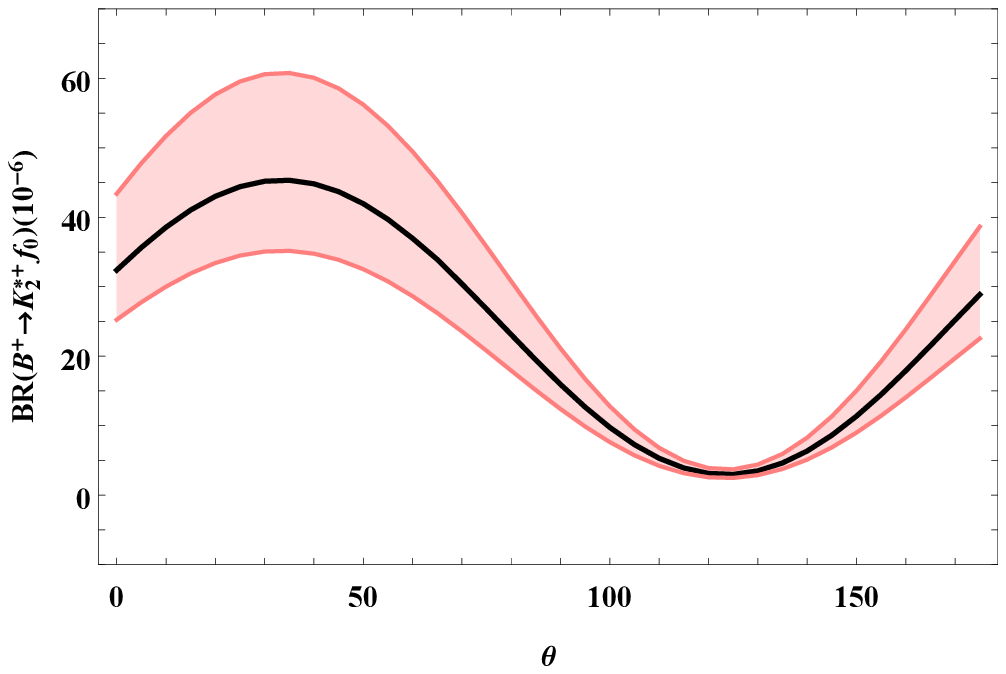}\\
\includegraphics[scale=0.65]{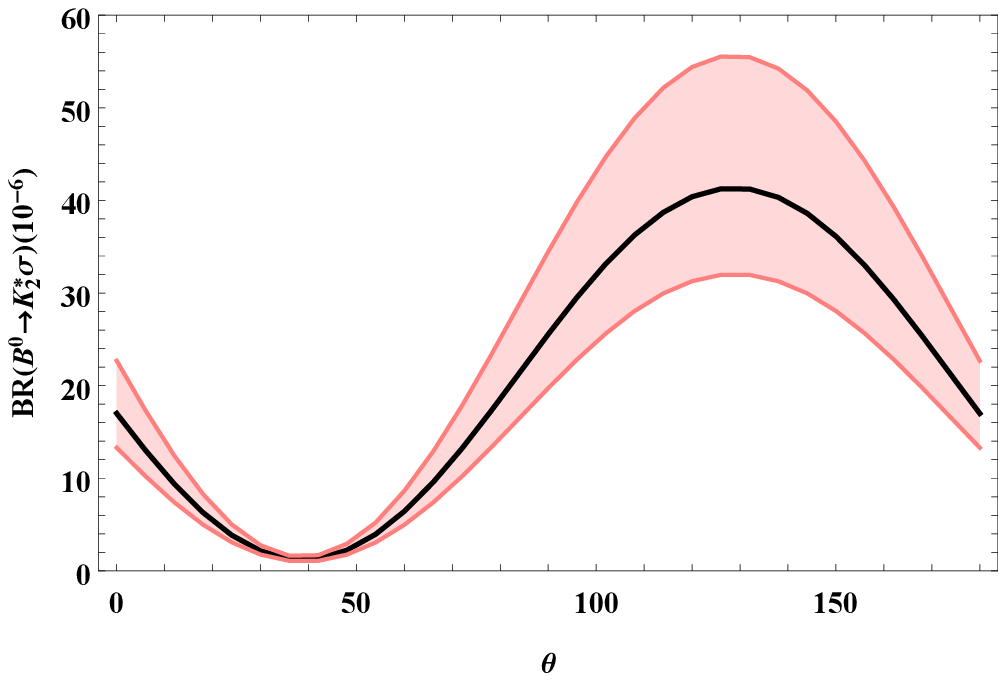}\,\,\,\,
\includegraphics[scale=0.65]{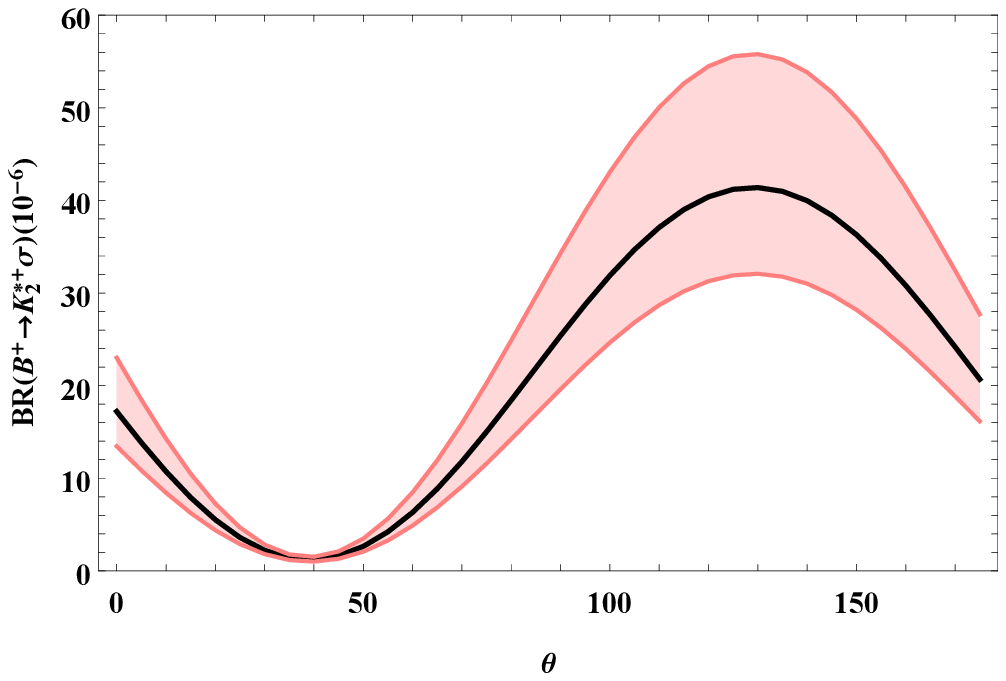}
\caption{The branching fractions of $B\to K_2^* f_0$ and $B\to K_2^* \sigma$ with variant of the mixing angle $\theta$. The black lines are the center values, and the horizontal (green) band is the experimental value}\label{Fig:3}
\end{center}
\end{figure}

By setting the mixing angle $\theta$ to be a free parameter, we plot the variation  of the branching fractions of decays  $B \to K_{0}^{*0}f_0(\sigma)$ and $ B \to K_2^{*0}f_0(\sigma)$ with the angle $\theta$ in Figure.\ref{Fig:2} and Figure.\ref{Fig:3}, respectively. We acknowledge that the uncertainty is the inevitable incident in theoretical evaluation.
In the present work, three kinds of errors are taken into account: the first errors are caused by the nonperturbative parameters, such as the initial and final mesons' wave functions aforementioned in Sec.~\ref{sec:function}, which are dominant in our calculation. Fortunately, these errors will be reduced with the improvement of the experiments and the update of the theoretical understanding. The second kind of errors come from the variations of the factorization scale ``$t$'' and  QCD scale $\Lambda_{\rm QCD}$, characterized by $0.8t\sim t\sim 1.2t$ and $\Lambda_{\rm QCD}=(0.25\pm 0.05)$ GeV. In fact, these errors reflect the contributions of the next-to-leading order radiative corrections and the next-to-leading order power corrections, since the complete next-to-leading order corrections in PQCD approach have not been accomplished \cite{Li:2012nk,Zhang:2014bsa} till now. The last errors are from the uncertainties of the unitary angle $\gamma$. We combine all uncertainties together and give the bounds as shown in the figures.

In ref. \cite{Lees:2011dq}, BaBar collaboration reported the first measurements of branching fractions of the $B^0\to K_{0,2}^{*0}f_0$ decays:
\begin{eqnarray}
\mathcal{B}(B^0\to K_0^{*0}f_0)=(2.7\pm0.7\pm0.6)\times 10^{-6}, \\
\mathcal{B}(B^0\to K_2^{*0} f_0)=(8.6\pm1.7\pm1.0)\times 10^{-6}, \label{babar}
\end{eqnarray}
which can be also found in the Figure.\ref{Fig:2} and Figure.\ref{Fig:3}.
Combining our theoretical results of $B^0\to K_{0,2}^{*0}f_0$ and the experimental measurements, we obtain the
mixing angle $\theta$  in the range of $[135^{\circ},155^{\circ}]$, which is consistent with the conclusions of the refs. \cite{Zhang:2016qvq, Li:2012sw, Cheng:2002ai}. In ref\cite{Zhang:2016qvq}, the mixing angle is constrained in the range $[135^{\circ}, 158^{\circ}]$ by studying the charmed $B$ decays $B_{(s)} \to \overline{D}^0 f_0$. The authors in \cite{Li:2012sw} obtained the mixing angle $\theta\sim146^{\circ}$ by analyzing the charmonium decays $B_s\to J/\psi f_0(\sigma)$. If the $f_0$ is composed entirely of $s\bar{s}$ component, which indicates  $\theta= 0^{\circ}$, the branching fractions of $B^0\to K_0^{*0}f_0$ and $B^0\to K_2^{*0}f_0$ are about $1.0\times10^{-4}$ and $3.0\times10^{-5}$, respectively, both of which are much larger than the data provided by the BaBar collaboration. When the mixing is taken in account and assuming the mixing angle less than $90^{\circ}$, we find that the contributions from the component $n\bar{n}=(u\bar{u}+d\bar{d})/\sqrt{2}$ and $s\bar{s}$ have the same sign. Furthermore, due to the constructive interference between the two different type amplitudes from the two components, the branching fractions would be enhanced and overshoot the upper limit of the experimental data, which implies that the acute angle $\theta$ is unfavored. Conversely, if the angle $\theta>90^{\circ}$, the branching fractions of $B^0\to K_{0(2)}^{*0}f_0$ will be suppressed by the cancellation between these two amplitudes from $n\bar{n}$ and $s\bar{s}$ components, and the theoretical predictions of PQCD approach will accommodate the experimental data well.

Now, taking the $\theta=145^\circ$ as a benchmark, we present our predictions of branching fractions as
\begin{eqnarray}
\begin{array}{ll}
\mathcal{B}(B^0 \to K_0^{*0}f_0)=(2.8^{+3.0}_{-1.7})\times 10^{-6},
&  \mathcal{B}(B^+ \to K_0^{*+}f_0)=(2.7^{+2.9}_{-1.7})\times 10^{-6}, \\
\mathcal{B}(B^0 \to K_0^{*0}\sigma)=(298.0^{+96.7}_{-74.3})\times 10^{-6},
&  \mathcal{B}(B^+ \to K_0^{*+}\sigma)=(299.7^{+80.9}_{-66.3})\times 10^{-6}, \\
\mathcal{B}(B^0 \to K_2^{*0}f_0)=(8.8^{+3.1}_{-1.7})\times 10^{-6},\,
&  \mathcal{B}(B^+ \to K_2^{*+}f_0)=(8.6^{+2.7}_{-1.7})\times 10^{-6}, \\
\mathcal{B}(B^0 \to K_2^{*0}\sigma)=(38.9^{+14.0}_{-9.4})\times 10^{-6},
&  \mathcal{B}(B^+ \to K_2^{*+}\sigma)=(38.4^{+13.3}_{-8.6})\times 10^{-6},
   \end{array}
\end{eqnarray}
which can be measured in the current experiments, such as LHCb and Belle-2.

\begin{figure}[!htb]
\begin{center}
\includegraphics[scale=0.65]{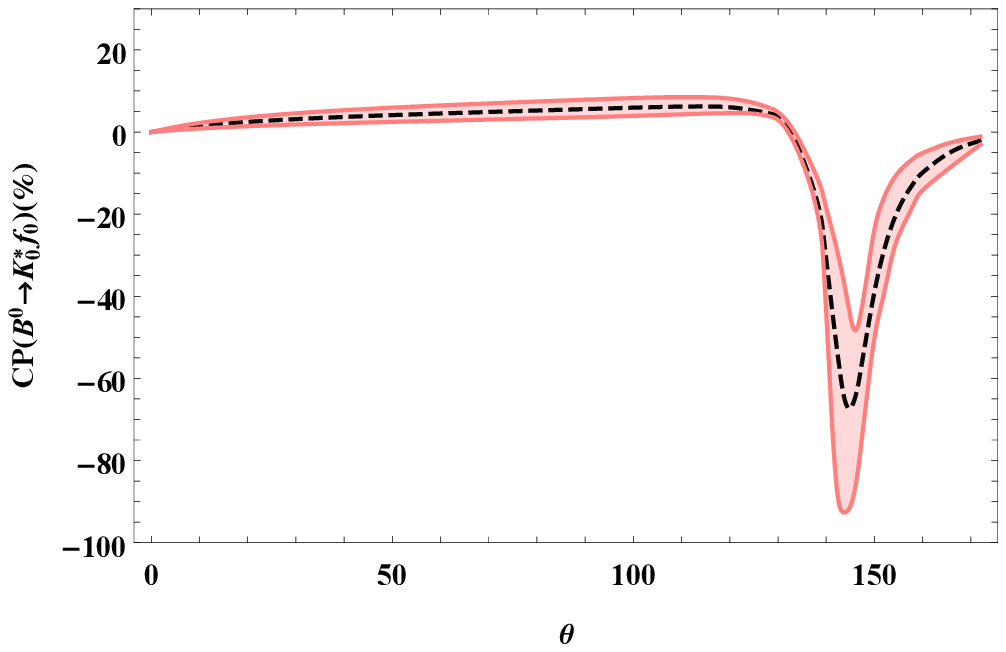}\,\,\,\,
\includegraphics[scale=0.65]{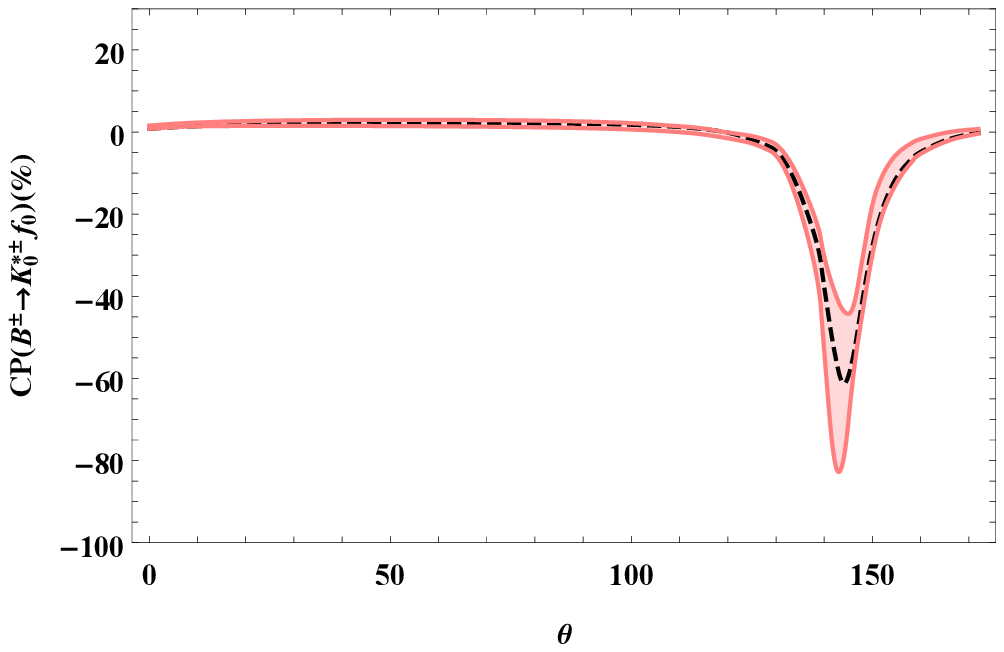}\\
\includegraphics[scale=0.65]{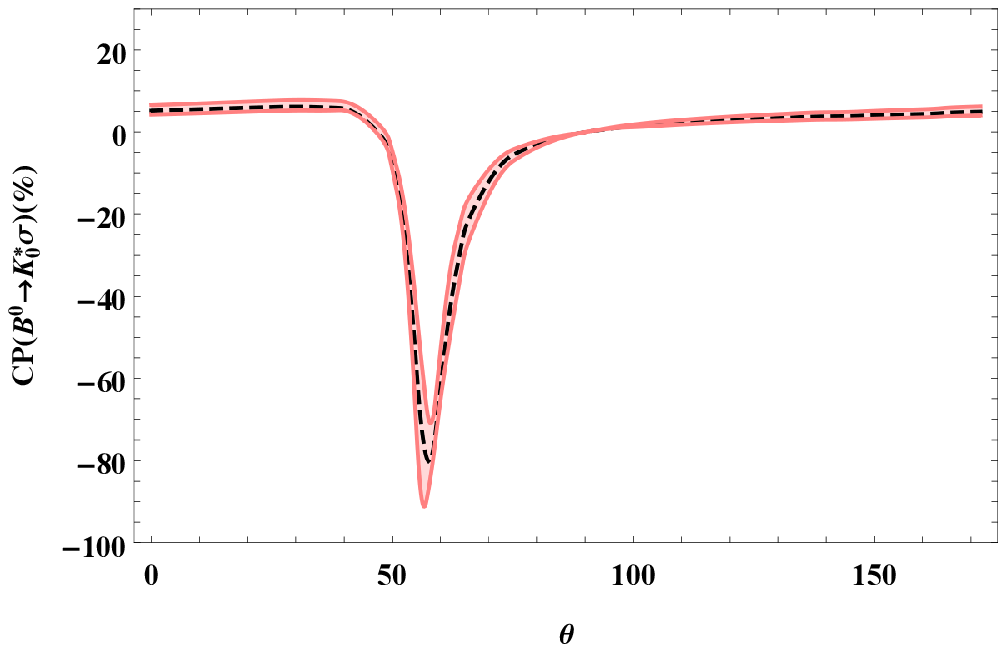}\,\,\,\,
\includegraphics[scale=0.65]{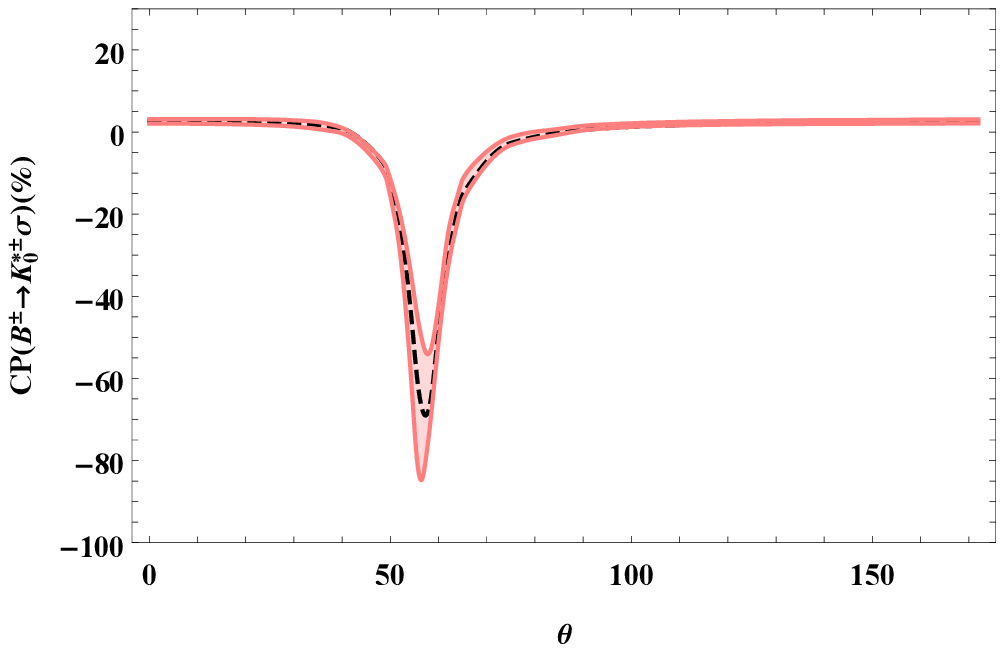}
\caption{The direct CP asymmetries of $B\to K_0^* f_0$ and $B\to K_2^* \sigma$ with variant of the mixing angle $\theta$.}\label{Fig:4}
\end{center}
\end{figure}

\begin{figure}[!htb]
\begin{center}
\includegraphics[scale=0.65]{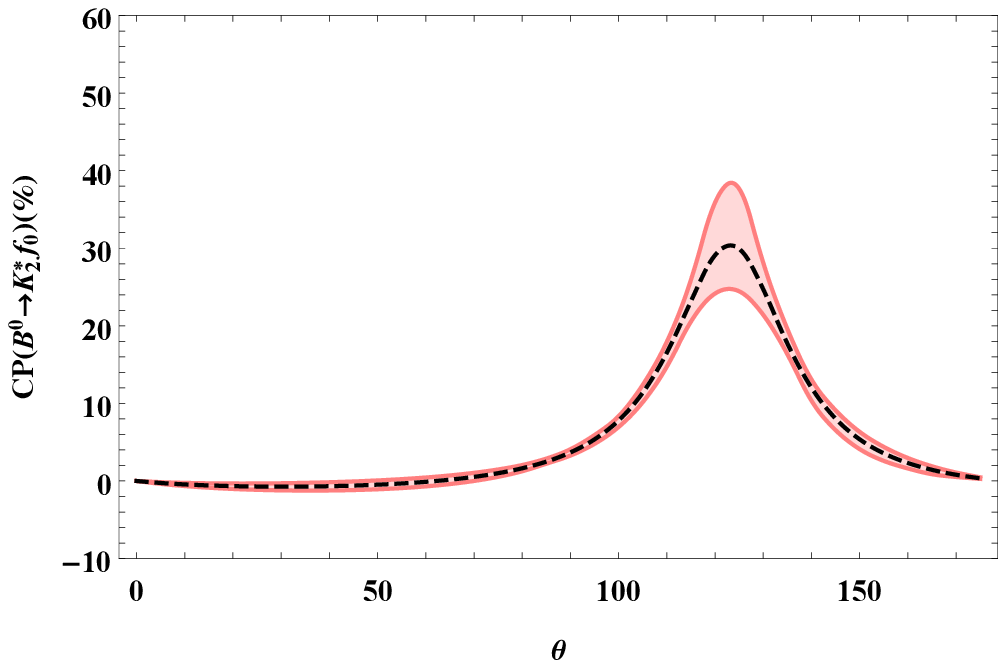}\,\,\,\,
\includegraphics[scale=0.65]{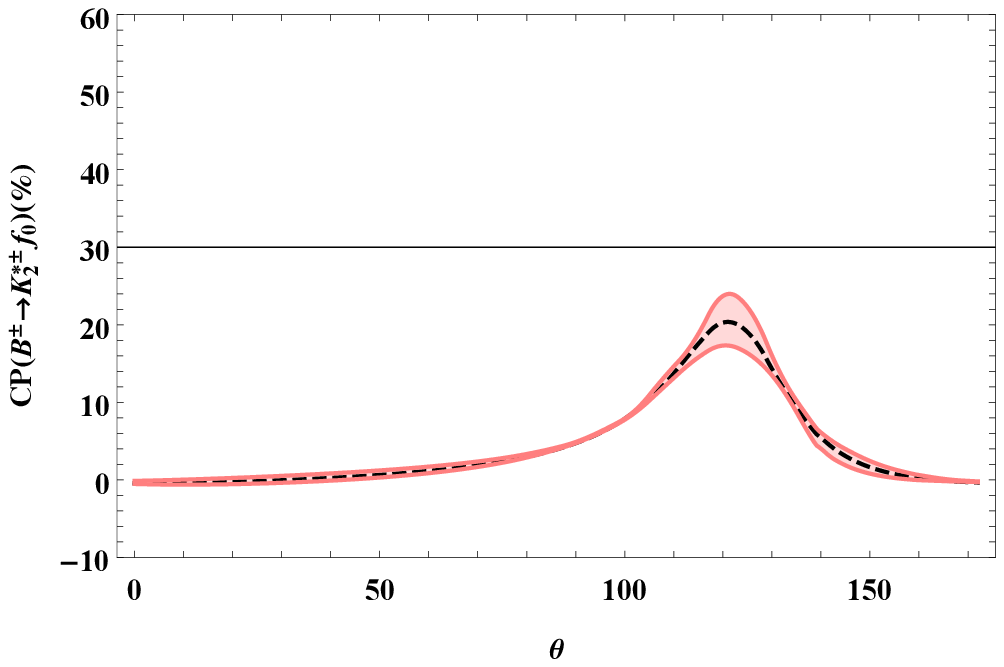}\\
\includegraphics[scale=0.65]{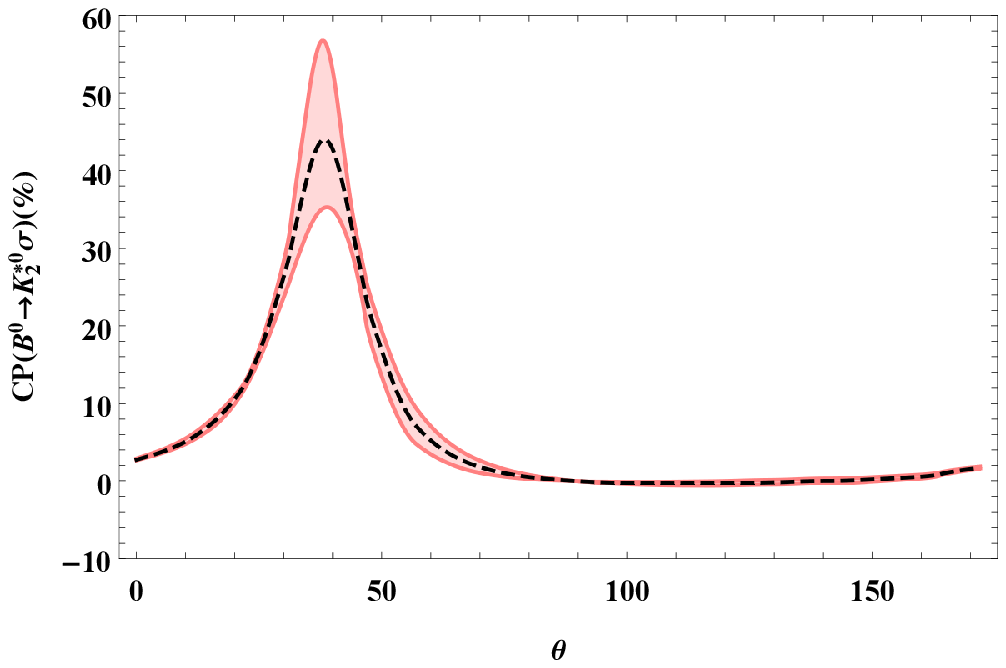}\,\,\,\,
\includegraphics[scale=0.65]{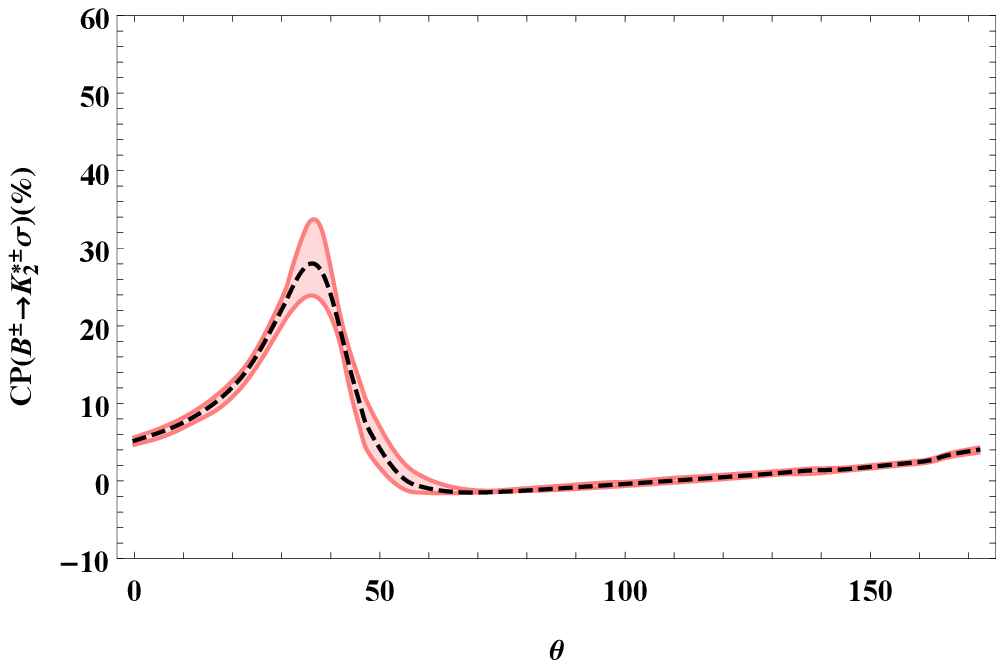}
\caption{The direct CP asymmetries of $B\to K_2^* f_0$ and $B\to K_2^* \sigma$ with variant of the mixing angle $\theta$.}\label{Fig:5}
\end{center}
\end{figure}

Lastly, we will discuss the relations between the direct CP asymmetries and the mixing angle. As we already known, both strong and weak phases are the necessary conditions for direct CP asymmetry. These decays concerned in this work are all governed by the $b\to s$ transition, and are dominated by the penguin operators, because the contributions from the tree operators are either forbidden or suppressed by small CKM matrix elements $|V_{us}V_{ub}|$. In the naive 2-quark model with the ideal mixing, the decay $B^0\to K_0^{*0}f_0$ and $B^0\to K_2^{*0}f_0$ are both induced by $b\to ss\bar s$ transition, which is a pure penguin process. In the Wolfenstein parameterization of CKM matrix, there is no weak phase in this transition, so the direct CP asymmetries of these two decays are zero. For $B^0\to K_0^{*0}\sigma$ and $B^0\to K_2^{*0}\sigma$ that are induced by $b\to sq\bar q$ ($q=u,d$), their direct CP asymmetries  decay are less than $5\%$, because $|V_{us}V_{ub}| \ll|V_{ts}V_{ub}|$. Since the mixing is supported by many experimental measurements and taken into account in this work, these considered decays receive three distinct types of contributions: the first one from the diagrams with emitted $K_{0(2)}^{*0(+)}$, the second one from the $f_0/\sigma$ emission with $q\bar{q}$ component and the last one from the $f_0/\sigma$ with $s\bar{s}$ component. Similar to the  branching fractions, these CP asymmetries are also related to the mixing angle $\theta$. We plot the CP asymmetries of these decays with the changes of the mixing angle $\theta$, as shown in Figure.\ref{Fig:4} and Figure.\ref{Fig:5}. When the mixing angle $\theta$ is involved, the $q\bar{q}$ component contributes to all concerned decays within the tree operators, which can cancel the penguin contributions from $s\bar{s}$ component when the mixing angle $\theta>90^{\circ}$. For instance, when the angle $\theta= 145^{\circ}$, the CP asymmetry of the $B^0\to K_0^{*0}f_0$ can be as large as $-68\%$. As for the $B^0\to K_{0(2)}^{*0}\sigma$ decays, the interference between $q\bar q$ and $s\bar s$ is contrary to corresponding decays with $f_0$. For the isospin asymmetry, we note that the interference for the considered $B^+$ decays are similar to the corresponding $B^0$ decays respectively, and $20\%$ differences can be attributed to the effects of tree operators in the annihilation diagrams, which can be found form the Figure.\ref{Fig:4} and Figure.\ref{Fig:5}. Because the direct CP asymmetry is a ratio, the theoretical uncertainties from the nonperturbative parameters will be cancelled, and the errors of these asymmetries will decrease, as illustrated in two figures. Therefore, if the two-quark structure will be confirmed, the CP asymmetries can also be used to determine the mixing angle $\theta$.

\section{Summary}
In this paper, it is the first time that  the $B^{0(+)}\to K_{0(2)}^{*}(1430)^{0(+)} f_0(980)(\sigma)$ decays were studied in the perturbative QCD approach under the two-quark assumption. Our theoretical results are hoped to shed light on the old puzzle about the inner structure of the scalar meson, especially the mixing angle of the $\sigma-f_0(980)$ system. For these decays, due to the charge conjugation invariance and the lorentz invariance, the factorizable emission diagrams are forbidden or suppressed heavily by the vector decay constants of scalar mesons, and the nonfactorizale diagrams and annihilation ones play the dominant roles. Moreover, for these considered penguin dominant decays, the penguin contributions from $n\bar{n}$ and $s\bar{s}$ components are at the same level. Thus the interferences are remarkable and affect the branching fractions and CP asymmetries significantly, which will provide us good platforms to determine the mixing angle. After the calculations, combining the experimental results of branching fractions, we find that, for the mixing angle, the range of $[135^{\circ}, 155^{\circ}]$ is favored. When the mixing angle $\theta=145^{\circ}$, the predicted branching ratios for $B^{0}\to K_{0(2)}^{*}(1430)^0 f_0(980)$ decays are in agreement with the experimental data well. The future measurements of CP asymmetries in LHCb and Belle-II can further test our results. Finally, we note that our calculation are only based on the two-quark assumption. The four-quark component or $K\overline{K}$ threshold effect that may be important components in $f_0(980)$  were not included, because the reliable nonperturbative input parameters are still absent and left for future study.
\section*{Acknowledgment}
This work was supported in part by the National Natural Science Foundation of China under the Grants No. 11705159, 11575151,11975195, 11765012, 11875033  and 11447032, and by the Natural Science Foundation of Shandong province under the Grant No. ZR2018JL001 and No. ZR2019JQ04. X. Liu is also supported by by the Qing Lan Project of Jiangsu Province under Grant No.~9212218405, and by the Research Fund of Jiangsu Normal University under Grant No.~HB2016004.

\bibliographystyle{bibstyle}
\bibliography{mybibfile}

\providecommand{\href}[2]{#2}\begingroup\raggedright\begin{thebibliography}{10}

\bibitem{Cheng:2009xz}
H.-Y. Cheng and J.~G. Smith, {\it {Charmless Hadronic B-Meson Decays}},  {\em
  Ann. Rev. Nucl. Part. Sci.} {\bf 59} (2009) 215--243,
  [\href{https://arxiv.org/abs/0901.4396}{{\tt arXiv:0901.4396}}].

\bibitem{Li:2018lxi}
Y.~Li and C.-D. Lu, {\it {Recent Anomalies in B Physics}},  {\em Sci. Bull.}
  {\bf 63} (2018) 267--269, [\href{https://arxiv.org/abs/1808.02990}{{\tt
  arXiv:1808.02990}}].

\bibitem{Abe:2002av}
{\bf Belle} Collaboration, K.~Abe et~al., {\it {Study of three-body charmless B
  decays}},  {\em Phys. Rev.} {\bf D65} (2002) 092005,
  [\href{https://arxiv.org/abs/hep-ex/0201007}{{\tt hep-ex/0201007}}].

\bibitem{Aubert:2003mi}
{\bf BaBar} Collaboration, B.~Aubert et~al., {\it {Measurements of the
  branching fractions of charged $B$ decays to $K^\pm \pi^\mp \pi^\pm$ final
  states}},  {\em Phys. Rev.} {\bf D70} (2004) 092001,
  [\href{https://arxiv.org/abs/hep-ex/0308065}{{\tt hep-ex/0308065}}].

\bibitem{Bondar:2004wr}
{\bf Belle} Collaboration, A.~Bondar, {\it {Dalitz analysis of $B^+\to K^+
  \pi^+ \pi^-$ and $B^+ \to K^+ K^+K^-$}},  in {\em {Proceedings, 32nd
  International Conference on High Energy Physics (ICHEP 2004): Beijing, China,
  August 16-22, 2004. Vol. 1+2}}, pp.~1125--1128, 2004.
\newblock \href{https://arxiv.org/abs/hep-ex/0411004}{{\tt hep-ex/0411004}}.

\bibitem{Garmash:2004wa}
{\bf Belle} Collaboration, A.~Garmash et~al., {\it {Dalitz analysis of the
  three-body charmless decays $B^+\to K^+ \pi^+ \pi^-$ and $B^+\to K^+ K^+
  K^-$}},  {\em Phys. Rev.} {\bf D71} (2005) 092003,
  [\href{https://arxiv.org/abs/hep-ex/0412066}{{\tt hep-ex/0412066}}].

\bibitem{Abe:2005ig}
{\bf Belle} Collaboration, K.~Abe, {\it {Search for direct CP violation in
  three-body charmless $B\pm \to K^\pm \pi^\pm \pi^\mp$ decay}},
  [\href{https://arxiv.org/abs/hep-ex/0509001}{{\tt hep-ex/0509001}}].

\bibitem{Abe:2005nya}
{\bf Belle} Collaboration, K.~Abe et~al., {\it {Study of $B^0\to \eta K^+
  \pi^-$ and $\eta \pi^+ \pi^-$}},
  [\href{https://arxiv.org/abs/hep-ex/0509003}{{\tt hep-ex/0509003}}].

\bibitem{Aubert:2004am}
{\bf BaBar} Collaboration, B.~Aubert et~al., {\it {Measurements of the
  branching fraction and CP-violation asymmetries in $B^0 \to f_0(980)
  K^0_S$}},  {\em Phys. Rev. Lett.} {\bf 94} (2005) 041802,
  [\href{https://arxiv.org/abs/hep-ex/0406040}{{\tt hep-ex/0406040}}].

\bibitem{Aubert:2004hs}
{\bf BaBar} Collaboration, B.~Aubert et~al., {\it {Search for B-meson decays to
  two-body final states with $a_0(980)$ mesons}},  {\em Phys. Rev.} {\bf D70}
  (2004) 111102, [\href{https://arxiv.org/abs/hep-ex/0407013}{{\tt
  hep-ex/0407013}}].

\bibitem{Aubert:2004xg}
{\bf BaBar} Collaboration, B.~Aubert et~al., {\it {Observation of B0 meson
  decays to $a^{+(1)}(1260) \pi^-$}},  in {\em {Proceedings, 32nd International
  Conference on High Energy Physics (ICHEP 2004): Beijing, China, August 16-22,
  2004. Vol. 1+2}}, 2004.
\newblock \href{https://arxiv.org/abs/hep-ex/0408021}{{\tt hep-ex/0408021}}.

\bibitem{Aubert:2004fn}
{\bf BaBar} Collaboration, B.~Aubert et~al., {\it {Amplitude analysis of $B^\pm
  \to \pi^\pm \pi^\mp \pi^\pm$ and $B^\pm \to K^\pm \pi^\mp \pi^\pm$}},  in
  {\em {Proceedings, 32nd International Conference on High Energy Physics
  (ICHEP 2004): Beijing, China, August 16-22, 2004. Vol. 1+2}}, 2004.
\newblock \href{https://arxiv.org/abs/hep-ex/0408032}{{\tt hep-ex/0408032}}.

\bibitem{Aubert:2004bt}
{\bf BaBar} Collaboration, B.~Aubert et~al., {\it {$B^0 \to K^{+} \pi^{-}
  \pi^0$ Dalitz plot analysis}},  in {\em {Proceedings, 32nd International
  Conference on High Energy Physics (ICHEP 2004): Beijing, China, August 16-22,
  2004. Vol. 1+2}}, 2004.
\newblock \href{https://arxiv.org/abs/hep-ex/0408073}{{\tt hep-ex/0408073}}.

\bibitem{Aubert:2005ce}
{\bf BaBar} Collaboration, B.~Aubert et~al., {\it {Dalitz-plot analysis of the
  decays $B^\pm \to K^\pm \pi^\mp \pi^\pm$}},  {\em Phys. Rev.} {\bf D72}
  (2005) 072003, [\href{https://arxiv.org/abs/hep-ex/0507004}{{\tt
  hep-ex/0507004}}]. [Erratum: Phys. Rev.D74,099903(2006)].

\bibitem{Aubert:2005sk}
{\bf BaBar} Collaboration, B.~Aubert et~al., {\it {An amplitude analysis of the
  decay $B^\pm \to \pi^\pm \pi^\pm \pi^\mp$}},  {\em Phys. Rev.} {\bf D72}
  (2005) 052002, [\href{https://arxiv.org/abs/hep-ex/0507025}{{\tt
  hep-ex/0507025}}].

\bibitem{Aubert:2005wb}
{\bf BaBar} Collaboration, B.~Aubert et~al., {\it {Measurements of neutral $B$
  decay branching fractions to $K^0_S \pi^+ \pi^-$ final states and the charge
  asymmetry of $B^0 \to K^{*+} \pi^-$}},  {\em Phys. Rev.} {\bf D73} (2006)
  031101, [\href{https://arxiv.org/abs/hep-ex/0508013}{{\tt hep-ex/0508013}}].

\bibitem{Jaffe:1976ig}
R.~L. Jaffe, {\it {Multi-Quark Hadrons. 1. The Phenomenology of (2 Quark 2
  anti-Quark) Mesons}},  {\em Phys. Rev.} {\bf D15} (1977) 267.

\bibitem{Alford:2000mm}
M.~G. Alford and R.~L. Jaffe, {\it {Insight into the scalar mesons from a
  lattice calculation}},  {\em Nucl. Phys.} {\bf B578} (2000) 367--382,
  [\href{https://arxiv.org/abs/hep-lat/0001023}{{\tt hep-lat/0001023}}].

\bibitem{Cheng:2005nb}
H.-Y. Cheng, C.-K. Chua, and K.-C. Yang, {\it {Charmless hadronic B decays
  involving scalar mesons: Implications to the nature of light scalar mesons}},
   {\em Phys. Rev.} {\bf D73} (2006) 014017,
  [\href{https://arxiv.org/abs/hep-ph/0508104}{{\tt hep-ph/0508104}}].

\bibitem{Tanabashi:2018oca}
{\bf Particle Data Group} Collaboration, M.~Tanabashi et~al., {\it {Review of
  Particle Physics}},  {\em Phys. Rev.} {\bf D98} (2018), no.~3 030001.

\bibitem{Anikina:2000tj}
M.~K. Anikina, A.~I. Golokhvastov, and J.~Lukstins, {\it {Dependence of the
  interferometric sizes of pion generation volume on sizes of their wave
  packet}},  {\em Phys. Atom. Nucl.} {\bf 65} (2002) 573--580. [Yad.
  Fiz.65,600(2002)].

\bibitem{Cheng:2002ai}
H.-Y. Cheng, {\it {Hadronic D decays involving scalar mesons}},  {\em Phys.
  Rev.} {\bf D67} (2003) 034024,
  [\href{https://arxiv.org/abs/hep-ph/0212117}{{\tt hep-ph/0212117}}].

\bibitem{Aloisio:2002bt}
{\bf KLOE} Collaboration, A.~Aloisio et~al., {\it {Study of the decay $\phi \to
  \pi^0 \pi^0 \gamma$ with the KLOE detector}},  {\em Phys. Lett.} {\bf B537}
  (2002) 21--27, [\href{https://arxiv.org/abs/hep-ex/0204013}{{\tt
  hep-ex/0204013}}].

\bibitem{Achasov:2000ym}
M.~N. Achasov et~al., {\it {The $\phi(1020)\to \pi^0 \pi^0 \gamma$ decay}},
  {\em Phys. Lett.} {\bf B485} (2000) 349--356,
  [\href{https://arxiv.org/abs/hep-ex/0005017}{{\tt hep-ex/0005017}}].

\bibitem{Akhmetshin:1999di}
{\bf CMD-2} Collaboration, R.~R. Akhmetshin et~al., {\it {Study of the $\phi$
  decays into $\pi^0 \pi^0 pi0 \gamma$ and $\eta \pi^0 \gamma$ final states}},
  {\em Phys. Lett.} {\bf B462} (1999) 380,
  [\href{https://arxiv.org/abs/hep-ex/9907006}{{\tt hep-ex/9907006}}].

\bibitem{Barberis:1999cq}
{\bf WA102} Collaboration, D.~Barberis et~al., {\it {A Coupled channel analysis
  of the centrally produced $K^+K^-$ and $\pi^+\pi^-$ final states in $pp$
  interactions at 450-GeV/c}},  {\em Phys. Lett.} {\bf B462} (1999) 462--470,
  [\href{https://arxiv.org/abs/hep-ex/9907055}{{\tt hep-ex/9907055}}].

\bibitem{Lees:2011dq}
{\bf BaBar} Collaboration, J.~P. Lees et~al., {\it {$B^0$ meson decays to
  $\rho^0 K^{*0}$, $f_0 K^{*0}$, and $\rho^-K^{*+}$, including higher $K^*$
  resonances}},  {\em Phys. Rev.} {\bf D85} (2012) 072005,
  [\href{https://arxiv.org/abs/1112.3896}{{\tt arXiv:1112.3896}}].

\bibitem{Giri:2006qk}
A.~K. Giri, B.~Mawlong, and R.~Mohanta, {\it {Probing new physics in $B\to f_0
  (980) K$ decays}},  {\em Phys. Rev.} {\bf D74} (2006) 114001,
  [\href{https://arxiv.org/abs/hep-ph/0608088}{{\tt hep-ph/0608088}}].

\bibitem{Cheng:2007st}
H.-Y. Cheng, C.-K. Chua, and K.-C. Yang, {\it {Charmless B decays to a scalar
  meson and a vector meson}},  {\em Phys. Rev.} {\bf D77} (2008) 014034,
  [\href{https://arxiv.org/abs/0705.3079}{{\tt arXiv:0705.3079}}].

\bibitem{Cheng:2010sn}
H.-Y. Cheng and C.-K. Chua, {\it {On Charmless $B\rightarrow K_{h}\eta^{(')}$
  Decays with $K_{h} = K$, $K^*$, $K_{0}^*(1430)$, $K_{2}^*(1430)$}},  {\em
  Phys. Rev.} {\bf D82} (2010) 034014,
  [\href{https://arxiv.org/abs/1005.1968}{{\tt arXiv:1005.1968}}].

\bibitem{Cheng:2010hn}
H.-Y. Cheng, Y.~Koike, and K.-C. Yang, {\it {Two-parton Light-cone Distribution
  Amplitudes of Tensor Mesons}},  {\em Phys. Rev.} {\bf D82} (2010) 054019,
  [\href{https://arxiv.org/abs/1007.3541}{{\tt arXiv:1007.3541}}].

\bibitem{Cheng:2010yd}
H.-Y. Cheng and K.-C. Yang, {\it {Charmless Hadronic B Decays into a Tensor
  Meson}},  {\em Phys. Rev.} {\bf D83} (2011) 034001,
  [\href{https://arxiv.org/abs/1010.3309}{{\tt arXiv:1010.3309}}].

\bibitem{Li:2011kw}
Y.~Li, X.-J. Fan, J.~Hua, and E.-L. Wang, {\it {Implications of Family
  Nonuniversal $Z^\prime$ Model on $B \to K_0^* \pi$ Decays}},  {\em Phys.
  Rev.} {\bf D85} (2012) 074010, [\href{https://arxiv.org/abs/1111.7153}{{\tt
  arXiv:1111.7153}}].

\bibitem{Li:2013aca}
Y.~Li, E.-L. Wang, and H.-Y. Zhang, {\it {Branching Fractions and $CP$
  Asymmetries of $B \to K^*_0(1430) \rho(\omega)$ and $B \to K^*_0(1430)\phi$
  Decays in the Family Nonuniversal $Z'$ Model}},  {\em Adv. High Energy Phys.}
  {\bf 2013} (2013) 175287, [\href{https://arxiv.org/abs/1206.4106}{{\tt
  arXiv:1206.4106}}].

\bibitem{Cheng:2013fba}
H.-Y. Cheng, C.-K. Chua, K.-C. Yang, and Z.-Q. Zhang, {\it {Revisiting
  charmless hadronic B decays to scalar mesons}},  {\em Phys. Rev.} {\bf D87}
  (2013), no.~11 114001, [\href{https://arxiv.org/abs/1303.4403}{{\tt
  arXiv:1303.4403}}].

\bibitem{Wang:2006ria}
W.~Wang, Y.-L. Shen, Y.~Li, and C.-D. Lu, {\it {Study of scalar mesons
  $f_0(980)$ and $f_0(1500)$ from $B\to f_0(980) K$ and $B \to f_0(1500) K$
  Decays}},  {\em Phys. Rev.} {\bf D74} (2006) 114010,
  [\href{https://arxiv.org/abs/hep-ph/0609082}{{\tt hep-ph/0609082}}].

\bibitem{Shen:2006ms}
Y.-L. Shen, W.~Wang, J.~Zhu, and C.-D. Lu, {\it {Study of K*0(1430) and a0(980)
  from B ---> K*0(1430) pi and B ---> a0(980)K Decays}},  {\em Eur. Phys. J.}
  {\bf C50} (2007) 877--887, [\href{https://arxiv.org/abs/hep-ph/0610380}{{\tt
  hep-ph/0610380}}].

\bibitem{Kim:2009dg}
C.~s. Kim, Y.~Li, and W.~Wang, {\it {Study of Decay Modes $B \to K_0^*(1430)
  \phi$}},  {\em Phys. Rev.} {\bf D81} (2010) 074014,
  [\href{https://arxiv.org/abs/0912.1718}{{\tt arXiv:0912.1718}}].

\bibitem{Liu:2009xm}
X.~Liu, Z.-Q. Zhang, and Z.-J. Xiao, {\it {B ---> K(0)*(1430) eta-prime decays
  in the pQCD approach}},  {\em Chin. Phys.} {\bf C34} (2010) 157--164,
  [\href{https://arxiv.org/abs/0904.1955}{{\tt arXiv:0904.1955}}].

\bibitem{Liu:2010zg}
X.~Liu and Z.-J. Xiao, {\it {B ---> K*0(1430) K decays in perturbative QCD
  approach}},  {\em Commun. Theor. Phys.} {\bf 53} (2010) 540,
  [\href{https://arxiv.org/abs/1004.0749}{{\tt arXiv:1004.0749}}].

\bibitem{Liu:2010kq}
X.~Liu and Z.-J. Xiao, {\it {Light scalar mesons and charmless hadronic $B_c
  \to SP, SV$ decays in the perturbative QCD approach}},  {\em Phys. Rev.} {\bf
  D82} (2010) 054029, [\href{https://arxiv.org/abs/1008.5201}{{\tt
  arXiv:1008.5201}}].

\bibitem{Wang:2010ni}
W.~Wang, {\it {B to tensor meson form factors in the perturbative QCD
  approach}},  {\em Phys. Rev.} {\bf D83} (2011) 014008,
  [\href{https://arxiv.org/abs/1008.5326}{{\tt arXiv:1008.5326}}].

\bibitem{Zou:2012td}
Z.-T. Zou, X.~Yu, and C.-D. Lu, {\it {Nonleptonic two-body charmless B decays
  involving a tensor meson in the Perturbative QCD Approach}},  {\em Phys.
  Rev.} {\bf D86} (2012) 094015, [\href{https://arxiv.org/abs/1203.4120}{{\tt
  arXiv:1203.4120}}].

\bibitem{Zou:2012sx}
Z.-T. Zou, X.~Yu, and C.-D. Lu, {\it {The $B(B_{s})\rightarrow
  D_{(s)}(\bar{D}_{(s)}) T$ and $D_{(s)}^{*}(\bar{D}_{(s)}^{*})T$ Decays in
  Perturbative QCD Approach}},  {\em Phys. Rev.} {\bf D86} (2012) 094001,
  [\href{https://arxiv.org/abs/1205.2971}{{\tt arXiv:1205.2971}}].

\bibitem{Zou:2012sy}
Z.-T. Zou, X.~Yu, and C.-D. Lu, {\it {The $B_c\rightarrow D^{(*)}T$ decays in
  perturbative QCD approach}},  {\em Phys. Rev.} {\bf D87} (2013) 074027,
  [\href{https://arxiv.org/abs/1208.4252}{{\tt arXiv:1208.4252}}].

\bibitem{Zou:2013wza}
Z.-T. Zou, R.~Zhou, and C.-D. Lu, {\it {Pure annihilation type decays $B^0 \to
  D^-_s K^{*+}_2$ and $B_s \to \bar{D} a_2$ in the perturbative QCD approach}},
   {\em Chin. Phys.} {\bf C37} (2013) 013103,
  [\href{https://arxiv.org/abs/1204.3144}{{\tt arXiv:1204.3144}}].

\bibitem{Liu:2013lka}
X.~Liu, Z.-J. Xiao, and Z.-T. Zou, {\it {Branching ratios and CP asymmetries of
  $B_{u,d,s}\to$ $K^*_0(1430) \overline K^*_0(1430)$ decays in the pQCD
  approach}},  {\em J. Phys.} {\bf G40} (2013) 025002.

\bibitem{Liu:2013cvx}
X.~Liu, Z.-J. Xiao, and Z.-T. Zou, {\it {Branching ratios and CP violations of
  $B\to K^*_0(1430)K^*$ decays in the perturbative QCD approach}},  {\em Phys.
  Rev.} {\bf D88} (2013), no.~9 094003,
  [\href{https://arxiv.org/abs/1309.7256}{{\tt arXiv:1309.7256}}].

\bibitem{Zou:2016yhb}
Z.-T. Zou, Y.~Li, and X.~Liu, {\it {Two-body charmed $B_s$ decays involving a
  light scalar meson}},  {\em Phys. Rev.} {\bf D95} (2017), no.~1 016011,
  [\href{https://arxiv.org/abs/1609.06444}{{\tt arXiv:1609.06444}}].

\bibitem{Zou:2017iau}
Z.-T. Zou, Y.~Li, and X.~Liu, {\it {Cabibbo-Kobayashi-Maskawa-favored $B$
  decays to a scalar meson and a $D$ meson}},  {\em Eur. Phys. J.} {\bf C77}
  (2017), no.~12 870, [\href{https://arxiv.org/abs/1704.03967}{{\tt
  arXiv:1704.03967}}].

\bibitem{Zou:2017yxc}
Z.-T. Zou, Y.~Li, and X.~Liu, {\it {Study of $B_c \to DS$ decays in the
  perturbative QCD approach}},  {\em Phys. Rev.} {\bf D97} (2018), no.~5
  053005, [\href{https://arxiv.org/abs/1712.02239}{{\tt arXiv:1712.02239}}].

\bibitem{Liu:2017cwl}
X.~Liu, R.-H. Li, Z.-T. Zou, and Z.-J. Xiao, {\it {Nonleptonic charmless decays
  of $B_c\to TP, TV$ in the perturbative QCD approach}},  {\em Phys. Rev.} {\bf
  D96} (2017), no.~1 013005, [\href{https://arxiv.org/abs/1703.05982}{{\tt
  arXiv:1703.05982}}].

\bibitem{Liu:2019ymi}
X.~Liu, Z.-T. Zou, Y.~Li, and Z.-J. Xiao, {\it {Phenomenological studies on the
  $B_{d,s}^0 \to J/\psi f_0(500) [f_0(980)]$ decays}},  {\em Phys. Rev.} {\bf
  D100} (2019), no.~1 013006, [\href{https://arxiv.org/abs/1906.02489}{{\tt
  arXiv:1906.02489}}].

\bibitem{Su:2019vbu}
L.~Su, Z.~Jiang, and X.~Liu, {\it {Studies on the $B \to \kappa \bar \kappa$
  decays in the perturbative QCD approach}},  {\em J. Phys.} {\bf G46} (2019),
  no.~8 085003, [\href{https://arxiv.org/abs/1906.04438}{{\tt
  arXiv:1906.04438}}].

\bibitem{Chang:1996dw}
C.-H.~V. Chang and H.-n. Li, {\it {Three - scale factorization theorem and
  effective field theory}},  {\em Phys. Rev.} {\bf D55} (1997) 5577--5580,
  [\href{https://arxiv.org/abs/hep-ph/9607214}{{\tt hep-ph/9607214}}].

\bibitem{Yeh:1997rq}
T.-W. Yeh and H.-n. Li, {\it {Factorization theorems, effective field theory,
  and nonleptonic heavy meson decays}},  {\em Phys. Rev.} {\bf D56} (1997)
  1615--1631, [\href{https://arxiv.org/abs/hep-ph/9701233}{{\tt
  hep-ph/9701233}}].

\bibitem{Keum:2000ph}
Y.-Y. Keum, H.-n. Li, and A.~I. Sanda, {\it {Fat penguins and imaginary
  penguins in perturbative QCD}},  {\em Phys. Lett.} {\bf B504} (2001) 6--14,
  [\href{https://arxiv.org/abs/hep-ph/0004004}{{\tt hep-ph/0004004}}].

\bibitem{Li:2001ay}
H.-n. Li, {\it {Threshold resummation for exclusive B meson decays}},  {\em
  Phys. Rev.} {\bf D66} (2002) 094010,
  [\href{https://arxiv.org/abs/hep-ph/0102013}{{\tt hep-ph/0102013}}].

\bibitem{Keum:2000wi}
Y.~Y. Keum, H.-N. Li, and A.~I. Sanda, {\it {Penguin enhancement and $B \to K
  \pi$ decays in perturbative QCD}},  {\em Phys. Rev.} {\bf D63} (2001) 054008,
  [\href{https://arxiv.org/abs/hep-ph/0004173}{{\tt hep-ph/0004173}}].

\bibitem{Lu:2000em}
C.-D. Lu, K.~Ukai, and M.-Z. Yang, {\it {Branching ratio and CP violation of
  $B\to\pi\pi$decays in perturbative QCD approach}},  {\em Phys. Rev.} {\bf
  D63} (2001) 074009, [\href{https://arxiv.org/abs/hep-ph/0004213}{{\tt
  hep-ph/0004213}}].

\bibitem{Ali:2007ff}
A.~Ali, G.~Kramer, Y.~Li, C.-D. Lu, Y.-L. Shen, W.~Wang, and Y.-M. Wang, {\it
  {Charmless non-leptonic $B_s$ decays to $PP$, $PV$ and $VV$ final states in
  the pQCD approach}},  {\em Phys. Rev.} {\bf D76} (2007) 074018,
  [\href{https://arxiv.org/abs/hep-ph/0703162}{{\tt hep-ph/0703162}}].

\bibitem{Zou:2015iwa}
Z.-T. Zou, A.~Ali, C.-D. Lu, X.~Liu, and Y.~Li, {\it {Improved Estimates of The
  $B_{(s)}\to V V$ Decays in Perturbative QCD Approach}},  {\em Phys. Rev.}
  {\bf D91} (2015) 054033, [\href{https://arxiv.org/abs/1501.00784}{{\tt
  arXiv:1501.00784}}].

\bibitem{Lu:2006fr}
C.-D. Lu, Y.-M. Wang, and H.~Zou, {\it {Twist-3 distribution amplitudes of
  scalar mesons from QCD sum rules}},  {\em Phys. Rev.} {\bf D75} (2007)
  056001, [\href{https://arxiv.org/abs/hep-ph/0612210}{{\tt hep-ph/0612210}}].

\bibitem{Buchalla:1995vs}
G.~Buchalla, A.~J. Buras, and M.~E. Lautenbacher, {\it {Weak decays beyond
  leading logarithms}},  {\em Rev. Mod. Phys.} {\bf 68} (1996) 1125--1144,
  [\href{https://arxiv.org/abs/hep-ph/9512380}{{\tt hep-ph/9512380}}].

\bibitem{Ali:1998eb}
A.~Ali, G.~Kramer, and C.-D. Lu, {\it {Experimental tests of factorization in
  charmless nonleptonic two-body B decays}},  {\em Phys. Rev.} {\bf D58} (1998)
  094009, [\href{https://arxiv.org/abs/hep-ph/9804363}{{\tt hep-ph/9804363}}].

\bibitem{Wirbel:1985ji}
M.~Wirbel, B.~Stech, and M.~Bauer, {\it {Exclusive Semileptonic Decays of Heavy
  Mesons}},  {\em Z. Phys.} {\bf C29} (1985) 637.

\bibitem{Bauer:1986bm}
M.~Bauer, B.~Stech, and M.~Wirbel, {\it {Exclusive Nonleptonic Decays of D,
  D(s), and B Mesons}},  {\em Z. Phys.} {\bf C34} (1987) 103.

\bibitem{Li:2012nk}
H.-n. Li, Y.-L. Shen, and Y.-M. Wang, {\it {Next-to-leading-order corrections
  to $B \to \pi$ form factors in $k_T$ factorization}},  {\em Phys. Rev.} {\bf
  D85} (2012) 074004, [\href{https://arxiv.org/abs/1201.5066}{{\tt
  arXiv:1201.5066}}].

\bibitem{Zhang:2014bsa}
Y.-L. Zhang, X.-Y. Liu, Y.-Y. Fan, S.~Cheng, and Z.-J. Xiao, {\it {$B\to
  \pi\pi$ decays and effects of the next-to-leading order contributions}},
  {\em Phys. Rev.} {\bf D90} (2014), no.~1 014029,
  [\href{https://arxiv.org/abs/1405.7103}{{\tt arXiv:1405.7103}}].

\bibitem{Zhang:2016qvq}
Z.-Q. Zhang, S.-Y. Wang, and X.-K. Ma, {\it {Insight into $f_0(980)$ through
  the $B_{(s)}$ charmed decays}},  {\em Phys. Rev.} {\bf D93} (2016), no.~5
  054034, [\href{https://arxiv.org/abs/1601.04137}{{\tt arXiv:1601.04137}}].

\bibitem{Li:2012sw}
J.-W. Li, D.-S. Du, and C.-D. Lu, {\it {Determination of $f_0-\sigma$ mixing
  angle through $B_s^0 $$\to $$J/\Psi$$f_0(980)(\sigma)$ decays}},  {\em Eur.
  Phys. J.} {\bf C72} (2012) 2229, [\href{https://arxiv.org/abs/1212.5987}{{\tt
  arXiv:1212.5987}}].

\end{thebibliography}\endgroup
\end{document}